# Odd-to-Even-Layer Switch of Spontaneous Excitonic State in $TiSe_2$ Thin Films

Kaili Wang [1,†], Zhuowei Liu [1,†], Yuyang Mu [1,†], Zhujuan Li [1], Zhenjie Fan [1], Yu Gui [1], Xiaodong Qiu [1], Baigeng Wang[1, 2,], Rui Wang [1, 2, 3*], Yi Zhang [1, 2, 3*], Qi-Kun Xue [4, 5, 6*]

[1] *National Laboratory of Solid State Microstructure, School of Physics, Nanjing University, Nanjing 210093, China*

[2] *Jiangsu Physical Science Research Center, Nanjing University, Nanjing 210093, China*

[3] *Hefei National Laboratory, Hefei 230088, China*

[4] *Quantum Science Center of Guangdong-Hong Kong-Macao Greater Bay Area, Shenzhen 518045, China*

[5] *State Key Laboratory of Low Dimensional Quantum Physics and Department of Physics, Tsinghua University, Beijing 100084, China*

[6] *Department of Physics and Guangdong Basic Research Center of Excellence for Quantum Science, Southern University of Science and Technology, Shenzhen 518055, China*

*** Email:** rwang89@nju.edu.cn
zhangyi@nju.edu.cn
qkxue@mail.tsinghua.edu.cn

**† These authors contributed equally to this work**

**Abstract**

$TiSe_2$ is a candidate of high temperature excitonic condensates with spontaneous exciton formation in addition to its charge density wave (CDW). However, direct observation and manipulation of this electron-hole paring state formation is still lacking. Here, using varying-temperature angle-resolved photoemission spectroscopy (VT-ARPES), we observe a distinct odd-to-even-layer switch of spontaneous exciton state formation with the evident correlated band structure in $TiSe_2$ thin films. While the odd-layer $TiSe_2$ films show a sharp band structure in single-particle picture superimposed by CDW renormalization, in the even-layer $TiSe_2$ films, the full-filling valence band (VB) opens a distinct notch at the top, and the conductive band (CB) shows a significant extra signal extending above Fermi level with spectra broadening. These features indicate that the extra holes (notched VB) and electrons (broadening CB above Fermi level) form spontaneous excitons in even-layer $TiSe_2$. Based on a correlated electron-hole model extracted from the tight-binding description, we reproduce the characterized band structures with spontaneous exciton formation, where the interlayer inter-band coupling in $TiSe_2$ plays a key role, which leads to an odd-to-even-layer switch of the excitonic state formation. Our results provide direct evidences of the spontaneous exciton formation, promising an excitonic condensate as versatile platform for realizing Bose-Einstein condensate (BEC) and investigating novel correlated phenomena.

## Introduction

Exciton, an electron-hole pairing state with strong correlation, are widely observed in many transition metal dichalcogenides (TMDCs) and moiré system, but usually has limit lifetime and needs external excitation [1-4]. A stable excitonic ground state with spontaneous exciton formation and condensation can exhibit intriguing correlation phenomena such as Bose-Einstein condensate (BEC), superfluid, and perfect Coulomb drag [5-10]. The interlayer excitonic condensates have been realized in several heterobilayer systems [6,8,10-13]. However, an intrinsic single material that possesses excitonic condensation state, namely excitonic insulator (EI), is still lacking [5,9,14]. Meanwhile, a solid spectroscopic characterization of the excitonic band structure is also crucial for understanding the formation of excitonic condensates.

$TiSe_2$, in additional to its $2 \times 2 \times 2$ charge density wave (CDW) order [15-21], is predicted to be an EI candidate by many theoretical works [22-24]. The experimental photoemission and momentum-resolved electron energy-loss spectroscopy results taken on the bulk $TiSe_2$ show signatures of exciton condensate phase [25-27]. The scanning tunneling spectroscopy (STS) and quasiparticle interference (QPI) results unveil the electron-exciton coupling in bilayer $TiSe_2$ thin film [28]. However, there are still debates on whether the exciton condensate state can exist in $TiSe_2$, especially in monolayer limit [24,29]. Directly investigating its correlated electronic structure under this excitonic state is a necessary way to understand the formation mechanics in this system.

In this letter, we realize the precisely control of the $TiSe_2$ film thickness from 1 layer (1L) to 4L in large area on $SrTiO_3(111)$ substrate via molecular beam epitaxy (MBE), promising the systemic investigations on the layer-dependent electronic structure evolution via in-situ angle-resolved photoemission spectroscopy (ARPES). We found that for the odd-layer $TiSe_2$ films (1L and 3L), the band structures exhibit a full-filling valence band (VB) at the Γ point and a conduction band (CB) crossing Fermi level at the M point, which can be simply considered as in single-particle picture superimposed by CDW renormalization. In contrast, for the even-layer $TiSe_2$ films (2L and 4L), the hole-type VB top open a notch with significant suppression of spectra weight, indicating the formation of extra holes. Meanwhile, an extra ARPES signal intensity appears above the Fermi level at the CB with significant spectra broadening, indicating that the depleted electrons from the VB top transfer to the unoccupied state at CB, and manifest strong correlation with the holes formed at the VB notch. With temperature increasing above ~250 K, the VB notch and CB broadening above Fermi level both disappear, and the band structure recover to a normal single-particle picture as that of odd-layer $TiSe_2$. Therefore, we conclude that the characterized notched VB and CB broadening above Fermi level in even-layer $TiSe_2$ are the solid evidences of the formation of electron-hole pairing of spontaneous excitons. Theoretically, starting from the tight-binding description of $TiSe_2$ films, we extract a correlated two band model with both intra and interlayer interactions. Hartree-Fock level mean-field calculations demonstrate formation of finite-momentum excitons on top of a CDW state. The strong tendency towards excitonic condensate generates the notch opening at VB. Moreover, the inter-band correlations further drive a finite quasi-particle lifetime with a significant spectral weight broadening, in agreement with our experimental observation. These results indicate that the interlayer coupling plays the key role in the odd-to-even-layer switch of the spontaneous excitonic state.

**Results and Discussion**

The unit-cell of 1T phase $TiSe_2$ is shown in Fig. 1(a) with $P\bar{3}m1$ space group, each layer (1L) of $TiSe_2$ consists of a Ti layer sandwiched by two Te layers. The substrate for the growth of $TiSe_2$ film is the prepared $SrTiO_3(111)$ with $SrO_3$ terminated surface [30,31]. Figs. 1(b)-(e) present the in-situ reflection high-energy electron diffraction (RHEED) patterns of the $SrTiO_3(111)$ substrate and grown 1L, 2L, and 4L $TiSe_2$ films, showing characterized $1 \times 1$ diffraction patterns. By measuring the spacings of diffraction stripes via the insetted intensity distribution curves (IDCs) in the RHEED patterns, we can obtain the in-plane lattice constant of the grown $TiSe_2$ films as $\boldsymbol{a} = 3.57 \pm 0.03$ Å for 1L, $3.54 \pm 0.03$ Å for 2L and 4L samples (the detailed calculation method is in the Supplementary Information part A). These values are same to the bulk $TiSe_2$ [32], and do not show obvious changing from 1L to 4L, indicating the van der Waals (vdW) epitaxy mode that can neglect lattice mismatch and interfacial stain effect. Fig. 1(f) presents the room-temperature scanning tunneling microscopic (RT-STM) image of the prepared $SrTiO_3(111)$ substrate, showing regular and atomically-flat terraces with characterized step heigh of 0.22 nm. The $SrTiO_3(111)$ substrate can provide enhanced interfacial interaction and result in a significant improvement of thickness controlling for the vdW epitaxy [31]. Fig. 1(g) is the RT-STM image of a 0.95L (95% coverage of 1L) $TiSe_2$ film, showing a uniform 1L thickness (0.65 nm) and absence of 2L island. The ~2L sample surface are mainly covered by 2L $TiSe_2$ with a few tiny 3L islands formed on surface [Fig. 1(h)]. For the ~4L sample, due to the weakening of interfacial interaction, a few small 5L islands form on the top of 4L $TiSe_2$, and a few 3L holes remain on the surface [Fig. 1(i)]. More STM images of 3L, 5L and 6L $TiSe_2$ films are in the Supplementary Information part B. In contrast, the $TiSe_2$ grown on graphene, HOPG, and $MoS_2$ substrates usually show a mixture of several thicknesses [18,28]. The precisely control of the $TiSe_2$ film thickness up to 4L by utilizing $SrTiO_3(111)$ substrate promises the symmetrically investigation on the layer-dependent electronic structures and exitonic states in $TiSe_2$ films.

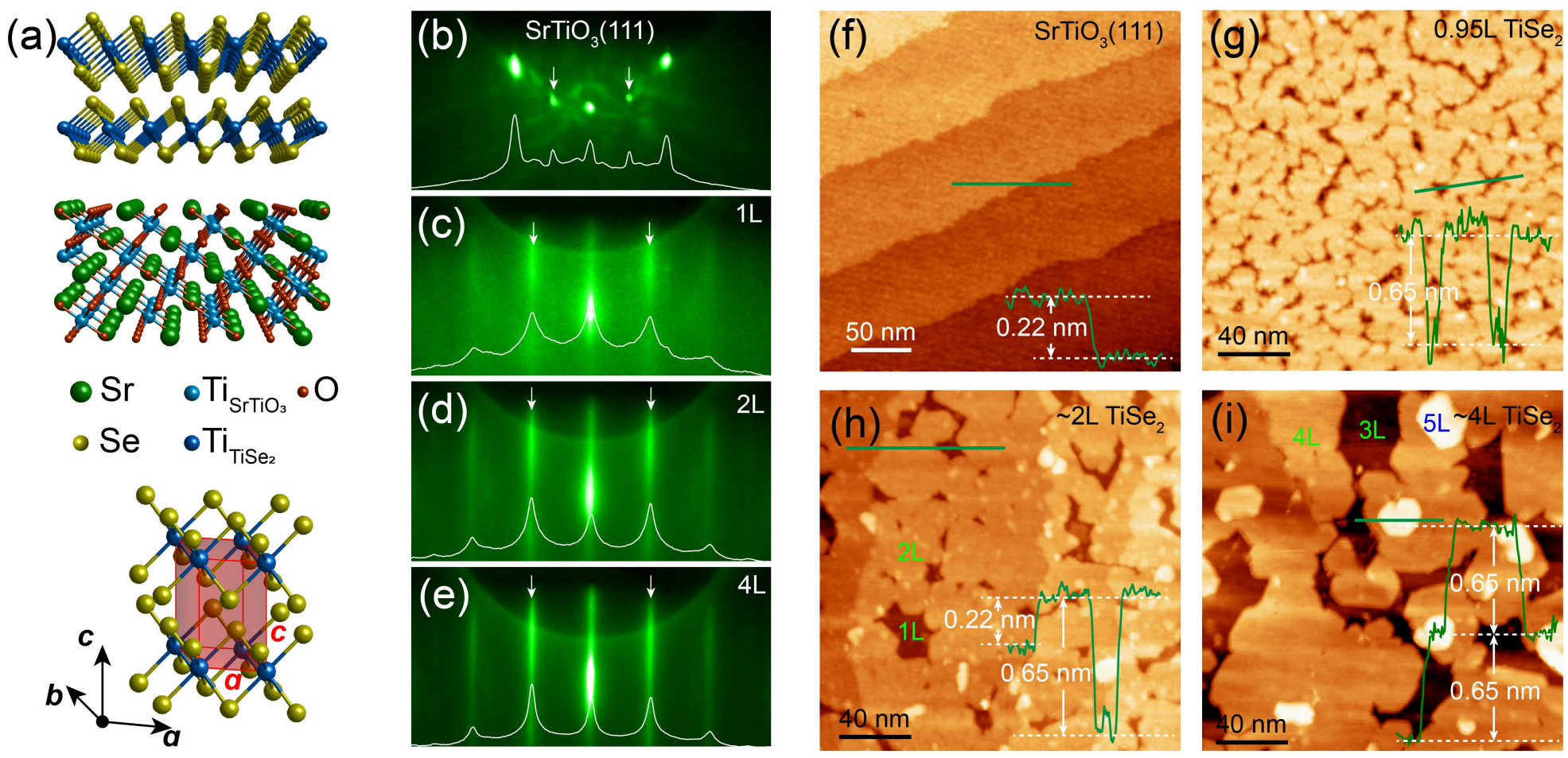

**Fig. 1. (a)** Lattice structure and unit-cell of $TiSe_2$ on $SrTiO_3(111)$ substrate. **(b)-(e)** RHEED patterns of (b) $SrTiO_3(111)$ substrate, (c) 1L, (d) 2L, (e) 4L $TiSe_2$ films, respectively. The white curves are the IDCs along horizontal direction. The white arrows indicate the $1 \times 1$ diffraction stripes of the substrate and films. **(f)-(i)** RT-STM images of (f) $SrTiO_3(111)$ substrate, (g) 0.95L, (h) ~2L, (i) ~4L $TiSe_2$ films, respectively. The insetted green curves are the height profiles along the green lines. Scanning parameter: $V_{bias}$ = 1.0 V, $I_t$ = 100 pA.

The in-situ ARPES measurements were conducted for investigating the layer-dependent electronic structures of $TiSe_2$ film. Fig. 2(a) presents the Fermi surface mapping of a 1L $TiSe_2$ film, showing elliptical electron pockets around the M point of $TiSe_2$ Brillouin zone (BZ). Figs. 2(b)-(e) are the ARPES spectra of 1L, 2L, 3L and 4L $TiSe_2$ films along the Γ-M direction. The 1L $TiSe_2$ shows two hole-type VBs at the Γ point below Fermi level, and a CB at the M point crossing Fermi level. From 2L to 4L, the top VB at the Γ point shows the characterized band folding phenomena to the M point, indicating the formation of $2 \times 2$ CDW order [16,17,28]. Although the band folding is absent in 1L $TiSe_2$, our later band structure calculation suggests that the CDW state can still exist in 1L limit. The absence of band folding is possibly due to the interface-induced suppression of band folding or modulation of CDW order [18,20,31]. Besides, the distinct evolution of VB structure from 1L to 4L can also evidence the precisely control of the film thickness [17].

Next, we scrutinize the top VB and CB evolutions in the zoom-in ARPES spectra shown in Figs. 2(f)-(m). For the odd-layer (1L and 3L) $TiSe_2$, the VB shows a typical hole-type parabolic dispersion with sharp and narrow spectral width in Figs. 2(f) & 2(h) (the yellow dashed curves are the parabolic fitting to the raw data). In contrast, for the even-layer (2L and 4L) $TiSe_2$, the spectra of the VB significantly broaden and open a distinct notch at the VB top, indicated by the magenta arrows in Figs. 2(g) & 2(i). This notch is more obvious in the corresponding second-derivate spectra. In intuitively and simplified understanding, this opening notch may suggest the non-full-filling of VB and the formation of extra free holes, compared to the full-filling VB of odd-layers. Based on the principle of

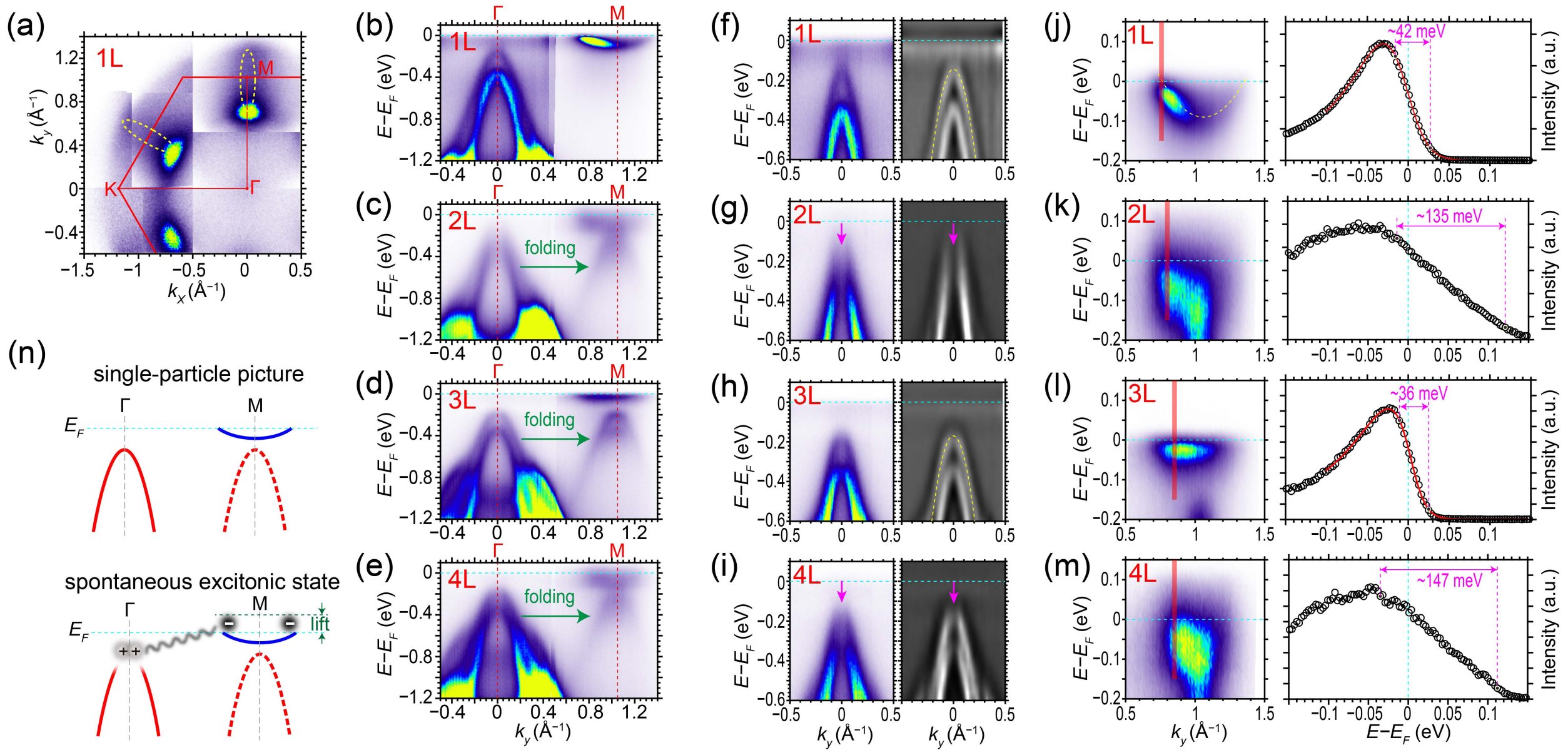

**Fig. 2. (a)** Fermi surface mapping of 1L $TiSe_2$ film. The red hexagon indicates the BZ of $TiSe_2$. **(b)-(e)** ARPES cut spectra of (b) 1L, (c) 2L, (d) 3L and (e) 4L $TiSe_2$ films along the Γ-M direction, respectively. **(f)-(i)** Zoom-in ARPES spectra (left panels) and the corresponding second-derivates (right panels) focused on the VB at the Γ point for (f) 1L, (g) 2L, (h) 3L and (i) 4L $TiSe_2$ films, respectively. The magenta arrows indicate the notch opening at the top of VB for 2L and 4L samples. **(j)-(m)** Zoom-in ARPES spectra focused on the CB at the M point for (j) 1L, (k) 2L, (l) 3L and (m) 4L $TiSe_2$ films, respectively. The right panels are the corresponding EDCs along the red lines in left panels. The magenta double-end arrows indicate the broadening width of 10%~90% intensity maximum at the Fermi level. **(n)** Schematics of the band structures based on the single-particle picture and simplified spontaneous excitonic state. All the ARPES spectra were taken at 10 K.

charge conservation, the depleted electrons from the notch of VB have to transfer to the unoccupied state of CB. Therefore, we plot the zoom-in spectra focused on the CB at the M point in Figs. 2(j)-(m). The CB of odd-layers shows a sharp Fermi cutting edge at the Fermi level, while for the even-layers, there is significant broadening of spectra weight and extra signal intensity above the Fermi level. To clearly illustrate this extra intensity, we extract the energy distribution curves (EDCs) along the red lines as shown in the right panels of Figs. 2(j)-(m). The magenta dashed lines and double-end arrows mark the 10%~90% of intensity maximum of EDCs for estimating the Fermi edge broadening width. For the odd-layers, the broadening width is as narrow as ~40 meV, generally agree with the total resolution ability of ARPES system. Meanwhile, the red Fermi function fitting curves march well with the raw data (detailed fitting method is in the Supplementary Information part C). In contrast, for the even-layers, the widths significantly broaden to ~140 meV, and extend far above the Fermi level [Figs. 2(k) & 2(m)], implying the existence of extra electrons above Fermi level at CB that are transferred from the VB notch, and the strong correlation effect of the CB electrons. More APRES data on the thicker $TiSe_2$ (5L and 6L) also exhibit such odd-to-even-layer switch (Supplementary Information part

D).

For better understanding the scenario of the spontaneous excitonic state in ARPES, we plot schematics in Fig. 2(n). For the odd-layer $TiSe_2$, its band structure can be considered simply as a normal single-particle picture superimposed by CDW renormalization [33], where a hole-type VB (drawn as the red curve) is at the Γ point below Fermi level, an electron-type CB (blue curve) is at the M point crossing Fermi level, and the folding VB (red dashed curve) at the M point caused by the $2 \times 2$ CDW order. The full-filling state of the VB indicates no free holes existed in the odd-layer $TiSe_2$, therefore no electron-hole pairing exciton can form in this normal single-particle picture. For the even-layer $TiSe_2$, the notched VB top signifies the existence of extra holes. The depleted electrons from VB notch transfer to the unoccupied state at CB above Fermi level, and pair with extra holes formed at the VB notch, forming the spontaneous excitonic state. The electron-hole pairing exciton is a composite boson, thus will result in the deviation of Fermi distribution and the broadening of Fermi edge as shown in Figs. 2(k) & 2(m). Besides, for the excitonic state, the extra intensity above the Fermi level can also be considered as the chemical potential lift of the CB, which is previously predicted by the BCS-like exciton theory [22,25].

To further evidence the spontaneous excitonic state in even-layer $TiSe_2$, we conducted the VT-ARPES measurements. Fig. 3(a) presents the VT-ARPES spectra and the corresponding second-derivates focused on the VB of a 2L $TiSe_2$ film. Significantly, when the temperature is above 250 K, the notch of the VB top closes, and recover as a sharp full-fulling hole-type parabolic band, indicating the recombination of extra holes. This transition can be more clearly distinguished from the EDCs along the Γ point as shown in Fig. 3(b). A distinct energy peak appears at around −0.15 eV (marked by the black arrows) when the temperature is above 250 K, indicating the closing of the notch at VB top.

Figures 3(c)&(d) compare the VT-ARPES spectra focused on the CB of 2L and 3L $TiSe_2$. The spectra of 3L $TiSe_2$ only show a normal Fermi distribution with thermal broadening. In contrast, for the 2L $TiSe_2$, when the temperature is above 250 K, the extra intensity above the Fermi level disappears, showing a normal Fermi distribution with thermal broadening, indicating the disappearance of the extra electrons accompanied by the closing of VB notch. To precisely determine the transition temperature and evaluate the Fermi edge broadening, we extract the EDCs along the $(\mathrm{M}-0.25\ \text{Å}^{-1})$ momentum position for 2L and along the $(\mathrm{M}-0.20\ \text{Å}^{-1})$ momentum position for 3L as shown in Figs. 3(e)&(f), where the CBs cross the Fermi level at the selected momentum positions. For the 3L $TiSe_2$,

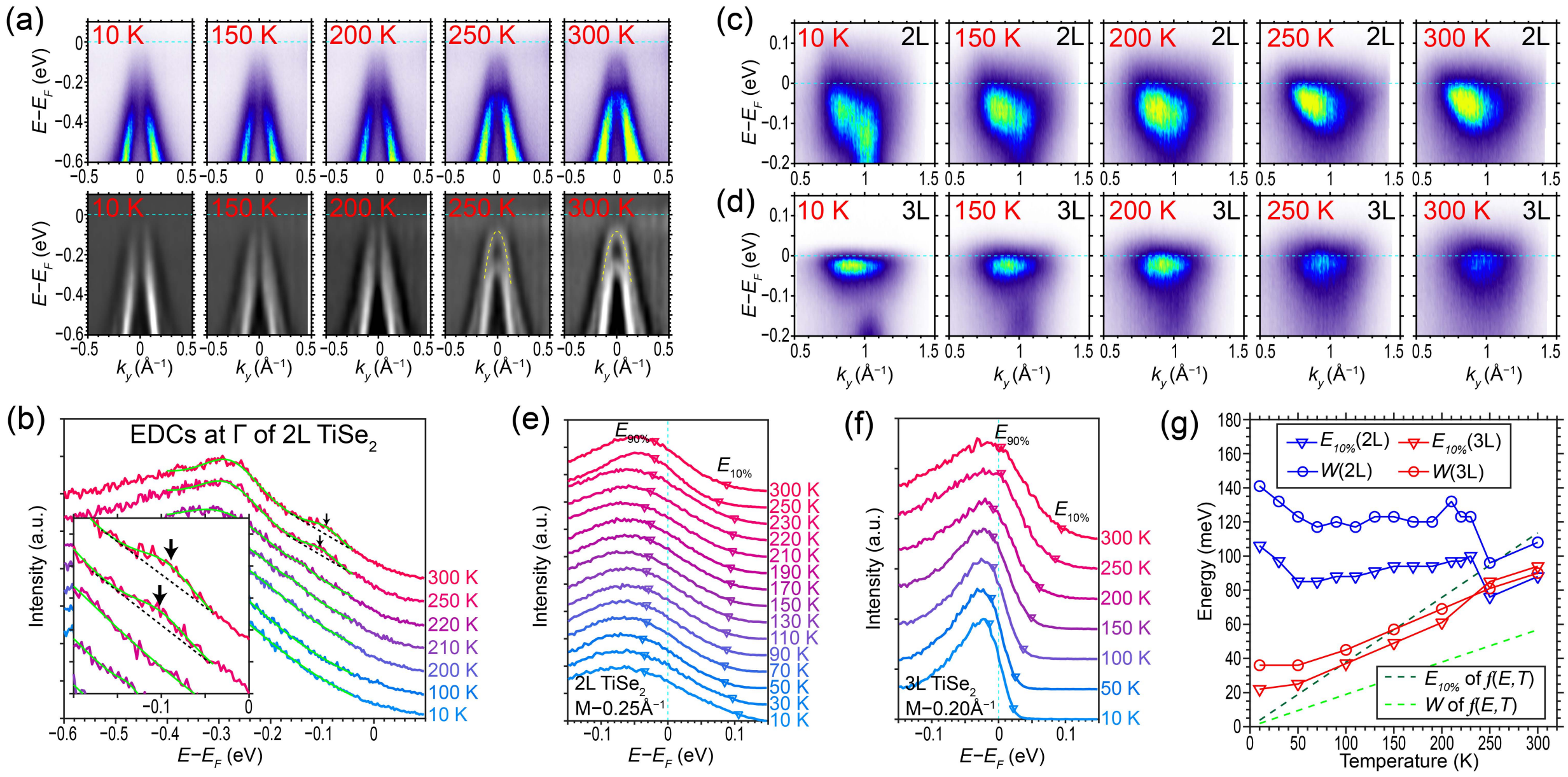

**Fig. 3 (a)** VT-ARPES spectra (upper panels) and the corresponding second-derivates (lower panels) focused on the VB at the Γ point of a 2L $TiSe_2$ film. **(b)** EDCs extracted from the VT-ARPES spectra along the Γ point of 2L $TiSe_2$. The green curves are the Lorentzian fittings to the raw data. Inset is the zoom-in EDCs. **(c)&(d)** VT-ARPES spectra focused on the CB at the M point of (c) 2L and (d) 3L $TiSe_2$ films, respectively. **(e)&(f)** EDCs extracted from the VT-ARPES spectra along the (M− 0.25 $Å^{-1}$) momentum position of 2L (e), and along the (M− 0.20 $Å^{-1}$) momentum position of 3L (f) $TiSe_2$ films, respectively. The triangles indicate the energy positions of 90% and 10% of intensity maximum ($E_{90\%}$ and $E_{10\%}$). **(g)** Temperature-dependent $E_{10\%}$ and $W = E_{10\%} - E_{90\%}$ values of 2L and 3L. The dark and light green curves are the $E_{10\%}$ and $W$ values of standard Fermi distribution function $f(E,T)$.

the broadening width keep increasing with rising temperature. In contrast, for the 2L $TiSe_2$, the broadening width remains unchanged from 10 K to 230 K, then suddenly shrinks at 250 K, and finally broadens from 250 K to 300 K. We further extract the energy positions of 10% of EDC maximum ($E_{10\%}$) and broadening widths $W = E_{10\%} - E_{90\%}$, then plot these values in Fig. 3(g). The $E_{10\%}$ and $W$ values of 3L agree with the standard Fermi distribution, only show small deviation below 50 K due to the resolution limit of ARPES system. In contrast, the $E_{10\%}$ and $W$ values of 2L show a distinct dip at the transition temperature of 250 K. The closing of VB notch and the recovery of Fermi edge at high temperature indicate that the recombination of extra holes and electrons caused by the thermal fluctuation, therefore evidencing spontaneous excitonic state in 2L $TiSe_2$ at low temperature. Besides, the final upturn of $E_{10\%}$ and $W$ values below 50 K may imply the possible BEC of the exciton, but may need further evidences [8]. More VT-ARPES spectra of 1L, 3L and 4L in the Supplementary Information part E can further evidence the spontaneous excitonic state in even-layer $TiSe_2$. Besides, the CDW transition with the disappearance of band folding are also observed in VT-ARPES (Supplementary Information part F).

To better understand the microscopic mechanisms of the spontaneous exciton formation and the observed odd-to-even-layer transitions, we studied the correlated electron-hole model for the single- and double-layer cases, with taking into account both the intra- and inter-layer interactions. We first calculated the band structure of 1L $TiSe_2$ at the single-particle level, based on a tight-binding description [34], As shown in Fig. 4(a) and the zoom-in data in Fig. 4(b), the VB top is located at the Γ point and the CB bottom is at the M point, consists with experiments. We then consider the intralayer interaction $V_{\mathrm{intra}}$. After projecting $V_{\mathrm{intra}}$ onto the lowest CB and VB near the Fermi energy, the monolayer system intends to form a CDW instability due to the Fermi surface nesting. The CDW slightly renormalizes the low-energy band structure, opening a small gap around the FS, as indicated by Fig. 4(c). In addition, the Fermi energy lies slightly above the bottom of the CB, as shown by Fig. 4(c). In principle, the formation of CDW mainly contributes to band structure renormalization without introducing the spectra weight broadening. Both the renormalized band structure and the Fermi energy quantitatively agree with the ARPES data of 1L $TiSe_2$. Although the interfacial effect from substrate could obscure the intensity of the folded VB around M point in 1L $TiSe_2$, the agreement in Fig. 4(c) indicates that the CDW state could be favored by the intralayer interaction in the $TiSe_2$ monolayer.

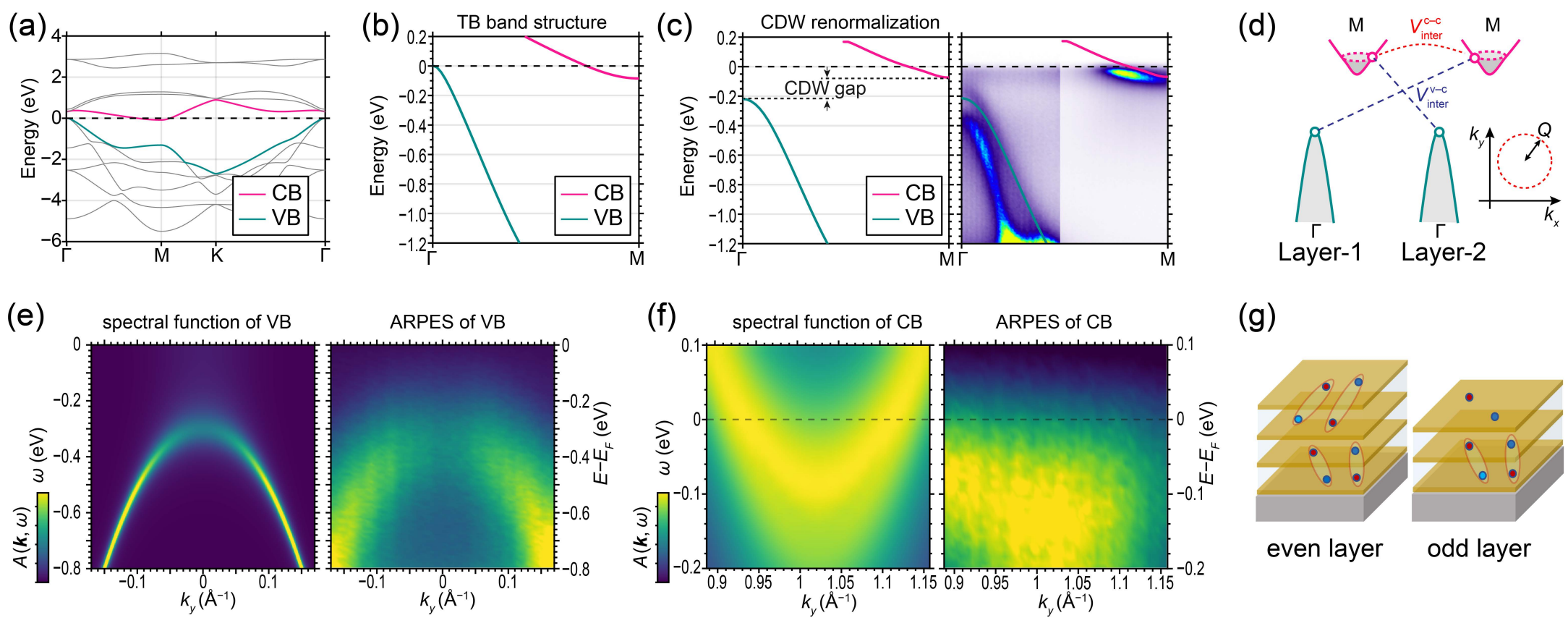

**Fig. 4 (a)** Tight-binding results on the band structure of 1L $TiSe_2$ at the single-particle level without considering correlations. **(b)** The zoom-in low-energy conduction and valence bands taken from (a). **(c)** The intralayer interaction drives a CDW instability in 1L $TiSe_2$, opening a small gap between the lowest conduction and valence band (left panel), and agreeing well the ARPES spectra (right panel). **(d)** The schematic plot of the correlation effects essential in the low-energy window for 2L $TiSe_2$. Two interaction channels, $V_{\mathrm{inter}}^{\mathrm{v-c}}$ and $V_{\mathrm{inter}}^{\mathrm{c-c}}$, are obtained after projection of the interlayer interaction into the lowest two bands. $V_{\mathrm{inter}}^{\mathrm{v-c}}$ drives pairing between virtual electron and hole excitations, leading to spontaneous excitons whose momenta intend to form a momentum loop shown by the inset. **(e)** Comparison of the calculated spectral function of the VB (left panel) with the ARPES spectra (right panel) of 2L $TiSe_2$ within same energy and momentum scales. **(f)** Comparison of the calculated spectral function of the CB (left panel) with the ARPES spectra (right panel) of 2L $TiSe_2$ within same energy and momentum scales. **(g)** Schematic plot illustrating the even-to-odd transition observed by experiment.

We then consider the correlation effect in 2L $TiSe_2$. Although the interlayer hopping is relatively weak in vdW layered materials, the interlayer interaction decays much slower, which is of the form $V_{\text{inter}}(q) = \frac{2\pi e^2}{\kappa q} e^{-d}$, where $\kappa$ is the dielectric constant and $d$ is the layer distance. We therefore propose a model describing two $TiSe_2$ monolayers — each favors a CDW — interacting with one another via $V_{\text{inter}}$ [See the Supplementary Information part G]. This interlayer correlation could host exotic excitonic states on top of the monolayer CDWs, beyond previous understandings of the material, as will be shown in the following.

After projecting the interlayer interaction to the lowest bands around the Fermi energy, the dominant low-energy interaction channels become clear, As illustrated by Fig. 4(d), there exists virtual excitations between layer-1 VB and layer-2 CB (or equivalent, between layer-2 VB and layer-1 CB), generating electron and hole pairs. When the valence-conduction band interaction $V_{\text{inter}}^{\text{v-c}}$ dominates over the CDW gap, the formation of electron-hole bound states, i.e., excitons, are expected. More detailed calculations verify that the finite-momentum excitonic instability will take place for sufficient $V_{\text{inter}}^{\text{v-c}}$, and the excitons intend to condense at momenta lying on $\boldsymbol{q} = |\boldsymbol{Q}|$ with $|\boldsymbol{Q}|$ being the Fermi wave vector of the CB, as plot in the inset of Fig. 4(d). Notably, the holes residing on the VB top are all hybridized to low-energy electrons around the Fermi surface. As a result, the hybridization strongly redistributes the quasi-particle weight. Our calculation shows that the spectral function is significantly suppressed around the top of the VB, leading to a notch structure as shown by Fig. 4(e). This is comparable with the ARPES spectra of 2L VB, which is plot in the same energy and momentum window and shown in the right panel. The good agreement of between the theory and experiment confirms the scenario of interlayer excitonic pairing channeling.

We now analyze additional correlation effects involving the CB. As indicated by Fig. 4(d), despite the VB-CB interaction discussed above, the projection of $V_{\text{inter}}$ generates another interaction channel, i.e., the interaction between the layer-1 and layer-2 CBs, $V_{\text{inter}}^{\text{c-c}}$. The electrons on the CB are renormalized by $V_{\text{inter}}^{\text{c-c}}$, resulting in a finite lifetime through the imaginary part of the self-energy. Moreover, since the bottom of the CB is filled, there remains unpaired CB electrons scattering with the formed excitons [35]. This emergent electron-exciton interaction is non-negligible for imbalanced case where the Fermi energy is away from charge neutral point [36]. As carefully analyzed in Supplementary Information part G, this further shortens the lifetime of CB electrons, leading to a strongly broadened spectral weight. A detailed computation leads to the spectra function of the CB shown in the left panel of Fig. 4(f), in accordance with the ARPES spectra of 2L CB shown in the right panel. Thus, the results in Figs. 4(e)&(f) altogether demonstrate the spontaneous formation of interlayer excitonic states in 2L $TiSe_2$, as well as their nontrivial correlation with the unpaired electrons.

The odd-to-even-layer switch observed in experiments can also be explained by the above interplay exciton mechanism. As indicated by Fig. 4(g), for generic even layer cases, the interaction $V_{\mathrm{inter}}$ can always drive spontaneous excitons between two neighboring layers, forming "paired bilayer" hosting interlayer excitonic states. However, for odd layer cases, the top layer would be left unpaired and remains as a weakly or even uncorrelated monolayer. Thus, the ARPES measurement—which mainly detects electronic states at the sample surface—produces distinct spectral information between the even and odd layer cases. This points to a novel odd-even effect in excitonic physics unreported before, which is mainly driven by the interlayer electron-hole correlation.

## Conclusions

In summary, the observed VB notch and extended CB broadening above Fermi level, together with the theoretical calculation, clearly evidence the spontaneous excitonic state in even-layer $TiSe_2$, providing a simple understanding of the spontaneous exciton formation in band structure. For the odd-layer $TiSe_2$, the lacking of an interlayer inter-band coupling fails to form the excitonic state due to the unpaired top layer, therefore exhibits normal band structure in single-particle picture superimposed on CDW state. The distinct band structure differences clearly evidence the novel odd-to-even-layer switch of spontaneous excitonic state formation in $TiSe_2$ thin films. In addition, the observed high transition temperature (250 K) of the excitonic state in $TiSe_2$ promises the further fundamental research and practical applications in optoelectronics. Regarding the final upturn of $E_{10\%}$ and $W$ values in 2L $TiSe_2$, the more evidences of possible BEC of the exciton may need in further research [8]. Our results not only clearly evidence the spontaneous excitonic state in even-layer $TiSe_2$, but also provide a distinct band structure feature for seeking and verifying excitonic state in other candidates.

## Methods

- **Sample growth**

The growth of $TiSe_2$ films was performed in an ultra-high vacuum MBE-STM-APRES combined system with base pressure of $2\times10^{-10}$ mbar. The 0.7% Nb-doped $SrTiO_3$(111) wafer was chosen as substrate, and was initially degassed at 600°C for 6h, followed by annealing at 920°C for 1h, forming an atomically flat $SrO_3$ terminated surface. The high-purity Ti and Se were evaporated from separate Knudsen cells at temperatures of 1380°C and 130°C, respectively, with the flux ratio of Ti : Se over 1:60. During the co-deposition of Ti and Se, the prepared $SrTiO_3$(111) was kept at 300°C. The growth rate is as low as 120 min per layer (min/L) for ensuring the precisely controlling of film thickness. The growth procedure was monitored by an in-situ RHEED system. The film morphology was characterized by an in-situ RT-STM.

- **ARPES characterizations**

The in-situ ARPES measurement was performed with a Scienta Omicron DA 30 analyzer and a monochromic helium lamp (He I mode, $h\nu$ = 21.218 eV). The light spot size is around ~1 $mm^2$. During the ARPES measurement, the samples were controlled from 10 K to 300 K by a closed-cycle cryogen and a heater.

- **Theoretical calculation**

The low-energy electronic structure of $TiSe_2$ was described using a tight-binding model, from which the valence and conduction bands closest to the Fermi level were retained. The intralayer interaction was treated at the mean-field level to describe the charge-density-wave (CDW)-induced reconstruction of the low-energy band structure. For bilayer $TiSe_2$, we further included the interlayer Coulomb interaction between neighboring layers. The dynamically screened interlayer interaction was evaluated within the random-phase approximation (RPA). After projection onto the low-energy valence and conduction bands, we retained the two dominant interaction channels: the interlayer conduction–conduction channel and the interlayer valence–conduction particle–hole channel.

The tendency toward interlayer exciton formation was characterized by the static inter-band particle–hole susceptibility. The excitonic instability is determined by the vanishing of the corresponding RPA denominator. In the simplified isotropic continuum model, the instability occurs on an approximately circular contour with $|\boldsymbol{Q}| \simeq k_F$, indicating that states near the valence-band maximum are predominantly coupled to conduction-band states near the Fermi surface. To account phenomenologically for the crystalline anisotropy of the real material, the finite-momentum excitonic state was modeled as an equal-amplitude Larkin–Ovchinnikov (LO) state containing the two opposite ordering vectors $\pm\boldsymbol{Q}$. For the spectral calculations, the ordering vectors were chosen along the $k_x$ direction, $\pm\boldsymbol{Q} = (\pm\boldsymbol{Q}, 0)$ .The single-particle spectra were calculated from the retarded Green's functions according to $A_\alpha(k,\omega) = -2\,\mathrm{Im}\,G_\alpha^R(k,\omega)$, where $\alpha$ denotes the valence or conduction band. The LO excitonic hybridization was explicitly included in the valence-band Green's function and produces a strong redistribution and depletion of spectral weight near the valence-band maximum.

The remaining conduction-band renormalization was evaluated from the interlayer conduction–conduction interaction. The corresponding retarded self-energy was separated into Hartree and Fock contributions. The dynamical Fock term provides a finite imaginary part and hence a finite quasiparticle lifetime and spectral broadening, whereas the uniform Hartree contribution gives only a momentum- and frequency-independent energy shift. In addition, because the system is away from charge neutrality, excess unpaired conduction-band electrons coexist with the interlayer excitons. Electron–exciton scattering therefore provides an additional intrinsic relaxation channel for both the conduction electrons and the excitons.

These finite lifetime effects were analyzed by computing the Feynman diagrams at one-loop order, contributed by the electron-exciton scattering (see Supplementary Information part G4), which leads to a non-negligible conduction-electron broadening $\eta_c$ and an exciton damping rate $\Gamma$. In the calculations presented in the main text, $\eta_c = 0.2$ eV and $\Gamma = 0.2$ eV are taken, reflecting the significant damping induced by the density imbalance [35]. The real part of the exciton self-energy mainly produces a smooth renormalization of the exciton dispersion, and was absorbed into the effective exciton dispersion. Further details of the tight-binding model, RPA susceptibility, self-energy calculations, excitonic instability, and LO spectral reconstruction are provided in Supplementary Information part G.

**Supplementary Information**

A. Calculation method of the lattice constant via RHEED pattern
B. RT-STM images of 3L, 5L and 6L $TiSe_2$ films
C. Fitting method for the Fermi edge
D. ARPES spectra of 5L and 6L $TiSe_2$ films
E. VT-ARPES spectra of 1L, 3L and 4L $TiSe_2$ films
F. VT-ARPES spectra on the VB folding in $TiSe_2$ films
G. Calculation method on the excitonic state in even-layer $TiSe_2$

**Funding**

This work is supported by the Quantum Science and Technology-National Science and Technology Major Project (No. 2021ZD0302803), the National Key R&D Program of China (Grant No. 2022YFA1403601, 2024YFA1410500), the Fundamental Research Program of Natural Science Foundation of Jiangsu Province (No. BK20243011), the National Natural Science Foundation of China (Grant No.12322402, No.12274206), the Natural Science Foundation of Jiangsu Province (No. BK20233001), the Scientific Research Innovation Capability Support Project for Young Faculty (Grant No. SRICSPYF-ZY2025164), and the Fundamental Research Funds for the Central Universities (Grant No. KG202501).

## Supplementary Information *for*
## *Odd-to-Even-Layer Switch of Spontaneous Excitonic State in $TiSe_2$ Thin Films*

Kaili Wang†, Zhuowei Liu†, Yuyang Mu†, Zhujuan Li, Zhenjie Fan, Yu Gui, Baigeng Wang, Rui Wang*, Yi Zhang*, Qi-Kun Xue*

*** Email:** rwang89@nju.edu.cn
zhangyi@nju.edu.cn
qkxue@mail.tsinghua.edu.cn

**† These authors contributed equally to this work**

**This Supplementary Information contains:**

## A. Calculation method of the lattice constant via RHEED pattern

To determine the lattice constant of the grown $TiSe_2$ film, we extract the intensity distribution curves (IDCs) from the blue dashed rectangle regions in RHEED patterns [Figs. S1(a)&(b)], as shown in Figs. S1(c)&(d). To precisely determine the $1 \times 1$ peaks positions, we perform the Lorentzian fitting to the raw data by using following equation:

$$I(x) = \text{Shirley}(x) + \frac{A}{1 + \frac{(x - x_0)^2}{w^2}} \tag{S1}$$

where the $\text{Shirley}(x)$ is the Shirley background of the fitting regions [1,2], $w$ is the width of the peak, $x_0$ is the peak position. The fitting results are plotted as the red curves shown in Fig. S1(c)&(d), then we can get the spacings of the $1 \times 1$ peaks as $L_{STO} = 62.3$ and $L_{TiSe_2} = 97.9$. By using the $SrTiO_3$ lattice constant $\boldsymbol{a}_{STO} = 3.905$Å as reference, we can calculate the in-plane lattice constant of $TiSe_2$ film as

$$\boldsymbol{a}_{TiSe_2} = \frac{L_{STO}}{L_{TiSe_2}} \cdot \sqrt{2}\boldsymbol{a}_{STO} \tag{S2}$$

The results are $\boldsymbol{a} = 3.57 \pm 0.03$ Å for 1L $TiSe_2$, and $\boldsymbol{a} = 3.54 \pm 0.03$ Å for the other thicknesses. The errors are estimated as 1 pixel error of the diffraction peaks positions.

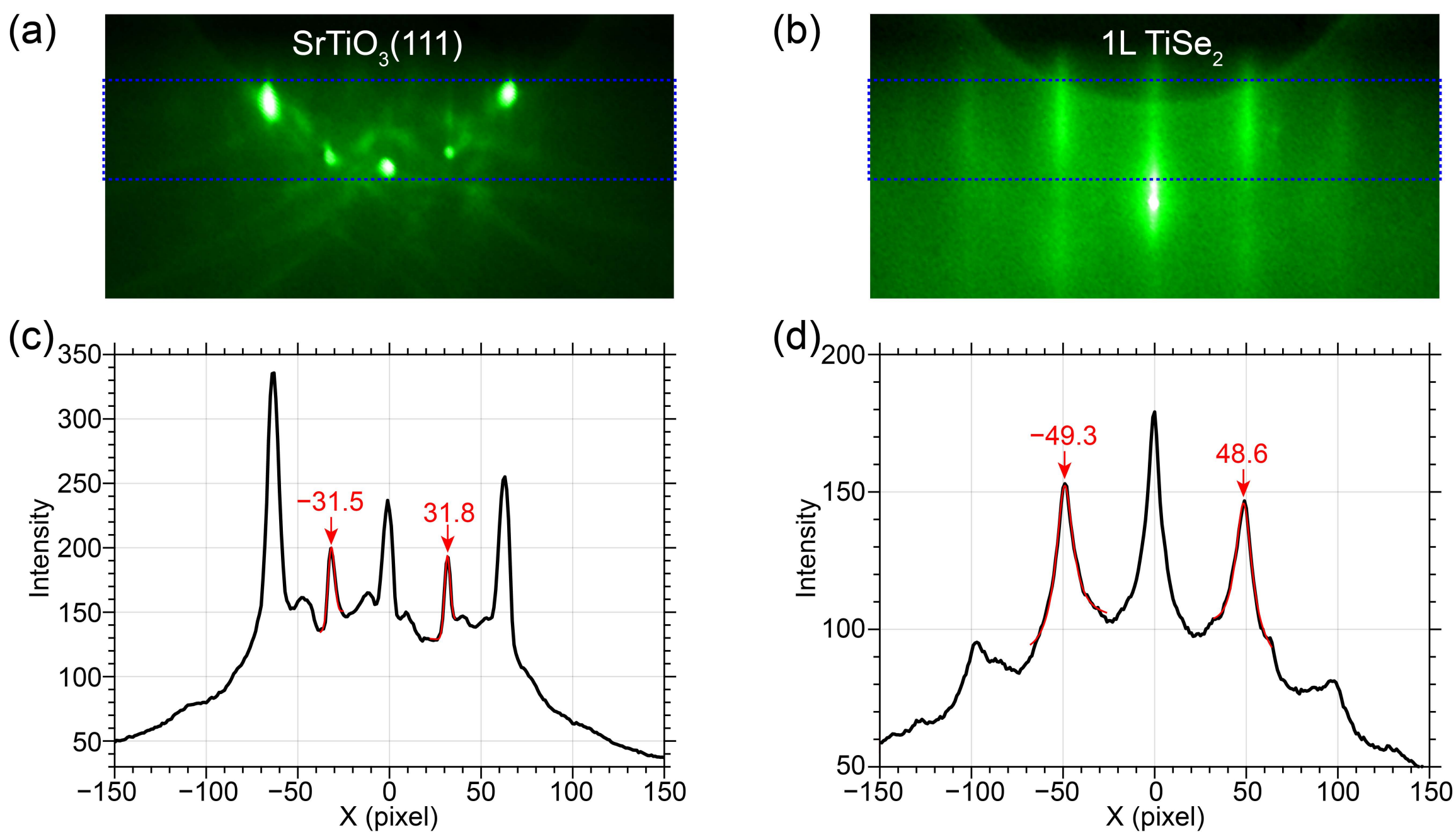

**Fig. S1. (a)&(b)** RHEED patterns of (a) $SrTiO_3(111)$ substrate and (b) 1L $TiSe_2$ film, respectively. **(c)&(d)** IDCs extracted from the blue dashed rectangle regions in RHEED patterns. The red curves are the Lorentzian fittings to the raw data.

## B. RT-STM images of 3L, 5L and 6LTiSe₂ films

Figure S2 presents the RT-STM images of 3L, 5L and 6L $TiSe_2$ films. For the 3L $TiSe_2$, the flatness is similar to the 2L $TiSe_2$ with very few 2L holes and 4L islands. For the 5L $TiSe_2$ film, the size of 4L holes and 6L islands increases, but the 5L region is still dominant. For the 6L $TiSe_2$ film, the size of 5L holes and 7L islands become larger more. The precisely control of the film thickness more than 6L becomes very difficult and challenging.

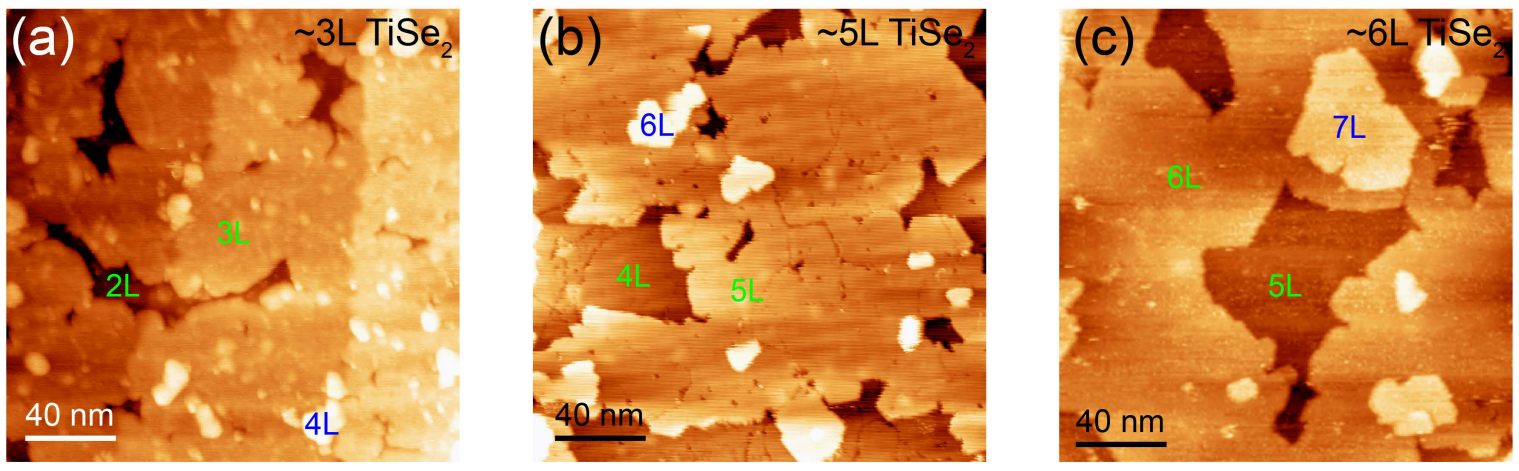

**Fig. S2 (a)-(c)** RT-STM images of (a) 3L, (b) 5L and (c) 6L $TiSe_2$ films, respectively.

## C. Fitting method for the Fermi edge

The 1L and 3L $TiSe_2$ show normal sharp Fermi edge as shown in the Figs. 2(j)&(l). To confirm this point, we using following equation to fit the EDCs at the Fermi level.

$$I(E) = \frac{A}{1 + \frac{(E - E_0)^2}{w^2}} \cdot \frac{1}{\exp\left(\frac{E - E_F}{k_B T}\right) + 1} \qquad \text{(S3)}$$

where the $\frac{A}{1+\frac{(E-E_0)^2}{w^2}}$ is the part of a Lorentzian peak, the $\frac{1}{\exp\left(\frac{E-E_F}{k_B T}\right)+1}$ is the standard Fermi distribution function. The fitting results are shown in the Fig. S3. The green curves of the Fermi function confirm the correct calibration of Fermi energy $E_F$ of the ARPES system. The red total fitting curves agree well with the raw data.

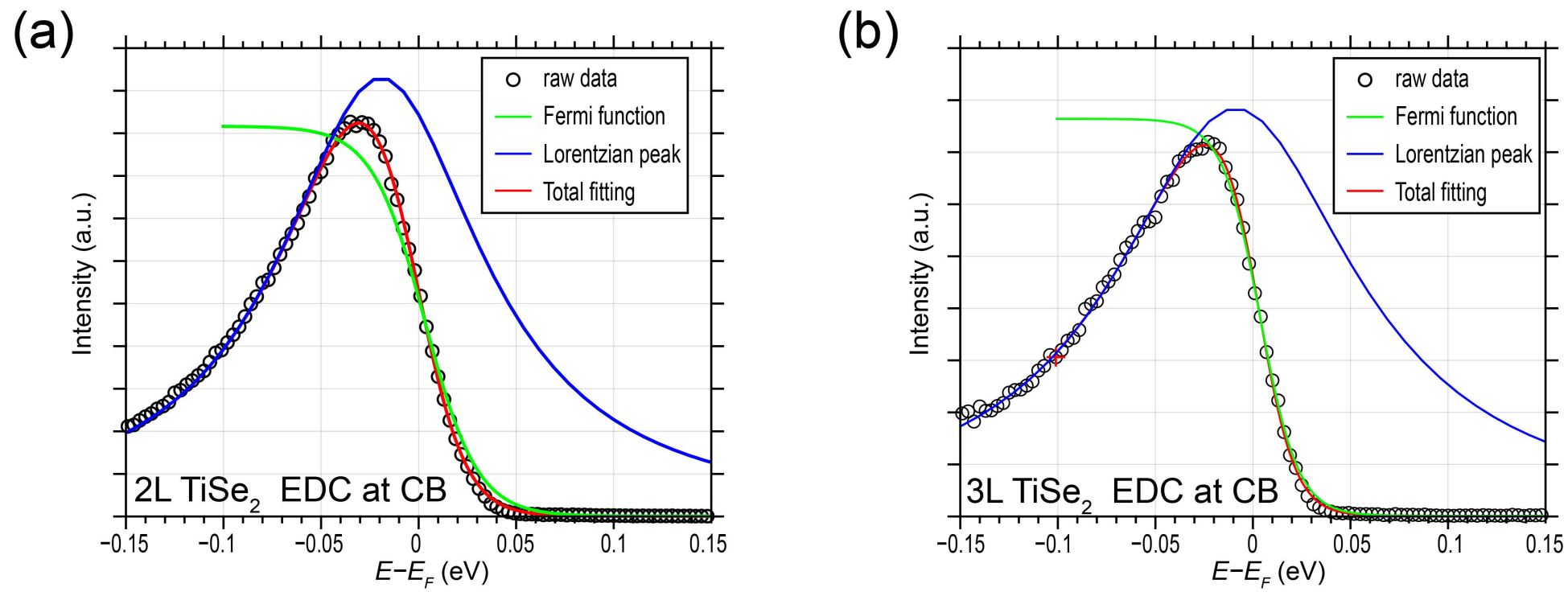

**Fig. S3 (a)&(b)** Fermi edge fittings to the CB EDCs of (a) 2L and (b) 3L $TiSe_2$, respectively.

## D. ARPES spectra of 5L and 6L $TiSe_2$ films

The band folding of VB is still observed in the 5L and 6L $TiSe_2$ films as shown in Figs. S4(a)&(b). In Fig. S4(c), the VB top of 5L show a weak notch opening due to the existence of the 4L holes and 6L islands as shown in Fig. S2(b). But for the 6L shown in Fig. S4(d), this notch becomes larger and more obvious. The odd-even-layer difference of the Fermi edge at CB is also observe in Figs. S4(e)&(f), the 5L film shows relatively narrower Fermi edge, while the 6L film shows a wider Fermi edge in the extent of 2L and 4L $TiSe_2$.

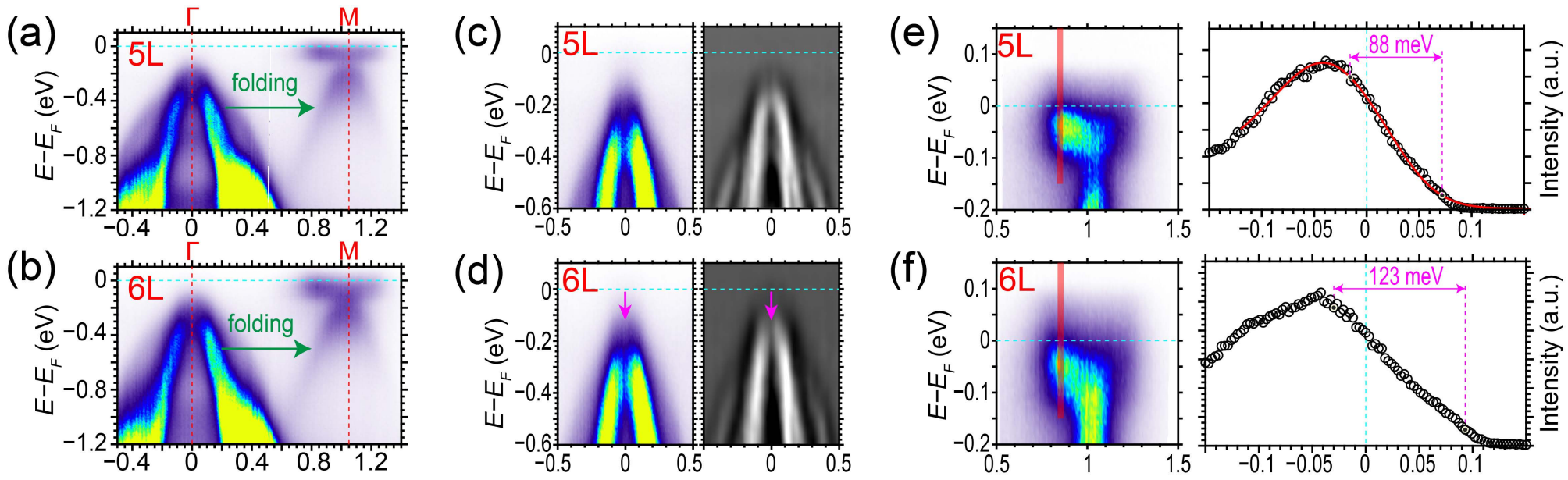

**Fig. S4. (a)&(b)** ARPES cut spectra of (a) 5L and (b) 6L $TiSe_2$ films along the Γ-M direction, respectively. **(c)&(d)** Zoom-in ARPES spectra (left panels) and the corresponding second-derivates (right panels) focused on the VB at the Γ point for (c) 5L and (d) 4L $TiSe_2$ films, respectively. The magenta arrows indicate the notch opening at the top of VB for 6L. **(e)&(f)** Zoom-in ARPES spectra focused on the CB at the M point for (e) 5L and (f) 6L $TiSe_2$ films, respectively. The right panels are the corresponding EDCs along the red lines in left panels. The magenta double-end arrows indicate the broadening width of 10%~90% intensity maximum at the Fermi level. All the ARPES spectra were taken at 10 K.

## E. VT-ARPES spectra of 1L, 3L and 4L $TiSe_2$ films

Figures S5(a)-(c) present the VT-ARPES spectra on the VB top of the 1L, 3L and 4L $TiSe_2$ films. For the 1L and 3L $TiSe_2$, the VB top shows sharp parabolic dispersion without notch opening, but only shift with increasing temperature. For the 1L $TiSe_2$, the VB top shifts downwards with the increasing temperature, possibly due to the temperature-dependent interfacial charge transfer effect of the $SrTiO_3(111)$ substrate [3-5]. For the 3L $TiSe_2$. the VB top shifts upwards with the increasing temperature, indicating the closing of CDW gap. Same to the 2L $TiSe_2$ in Fig. 3(a), the VB notch of 4L $TiSe_2$ also closes at 250 K as shown in Fig. S5(c).

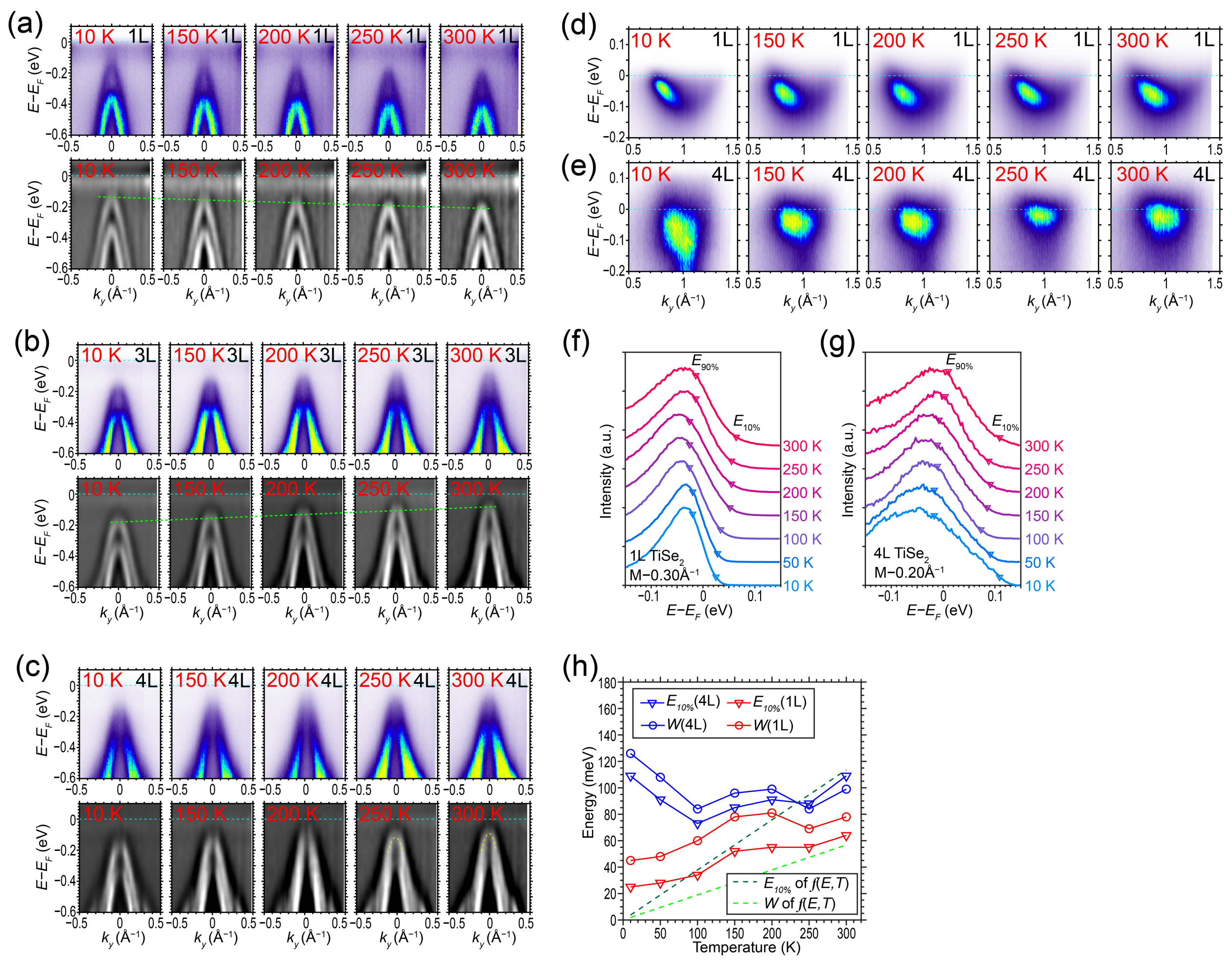

**Fig. S5. (a)-(c)** VT-ARPES spectra (upper panels) and the corresponding second-derivates (lower panels) focused on the VB at the Γ point of (a) 1L, (b) 2L and (c) 4L $TiSe_2$ films, respectively. **(c)&(d)** VT-ARPES spectra focused on the CB at the M point of (c) 2L and (d) 3L $TiSe_2$ films, respectively. **(d)&(e)** EDCs extracted from the VT-ARPES spectra along the (M – 0.30 $Å^{-1}$) momentum position of 1L (e), and along the (M – 0.20 $Å^{-1}$) momentum position of 4L (f) $TiSe_2$ films, respectively. The triangles indicate the energy positions of 90% and 10% of intensity maximum ($E_{90\%}$ and $E_{10\%}$). **(h)** Temperature-dependent $E_{10\%}$ and $W$=$E_{10\%}$−$E_{90\%}$ values of 1L and 4L. The dark and light green lines are the $E_{10\%}$ and $W$ values of standard Fermi distribution function $f(E,T)$.

Figures S5(d)&(e) present the VT-ARPES spectra on the CB of the 1L and 4L $TiSe_2$ films. The CB and Fermi edge of 1L $TiSe_2$ show normal broadening with increasing temperature. In contrast, for the 4L $TiSe_2$, the CB broadening significantly weakens from 10 K to 250 K, and enhances from 250 K to 300 K. We also plot temperature-dependent EDCs of 1L and 4L $TiSe_2$ in Figs. S5(f)&(g), and extract the $E_{10\%}$ and $W$ values as shown in Figs. S5(h). The varying-temperature behavior of these values also show the same distinct odd-even-layer difference as the 2L and 4L $TiSe_2$.

## F. VT-ARPES spectra on the VB folding in $TiSe_2$ films

Since the 1L $TiSe_2$ does not show the VB folding, therefore, we plot the VT-ARPES spectra on the VB folding of 2L, 3L and 4L $TiSe_2$ films in Fig. S6. The spectra intensity of the folded VB weakens with increasing temperature, and disappears at 200~250 K, indicating the CDW transition temperature. For the 2L and 4L $TiSe_2$, due the broadening VB and CB resulted by the exciton, we can not obtain the CDW gap. For the 3L $TiSe_2$, the CDW gap decreases with increasing temperature [6,7].

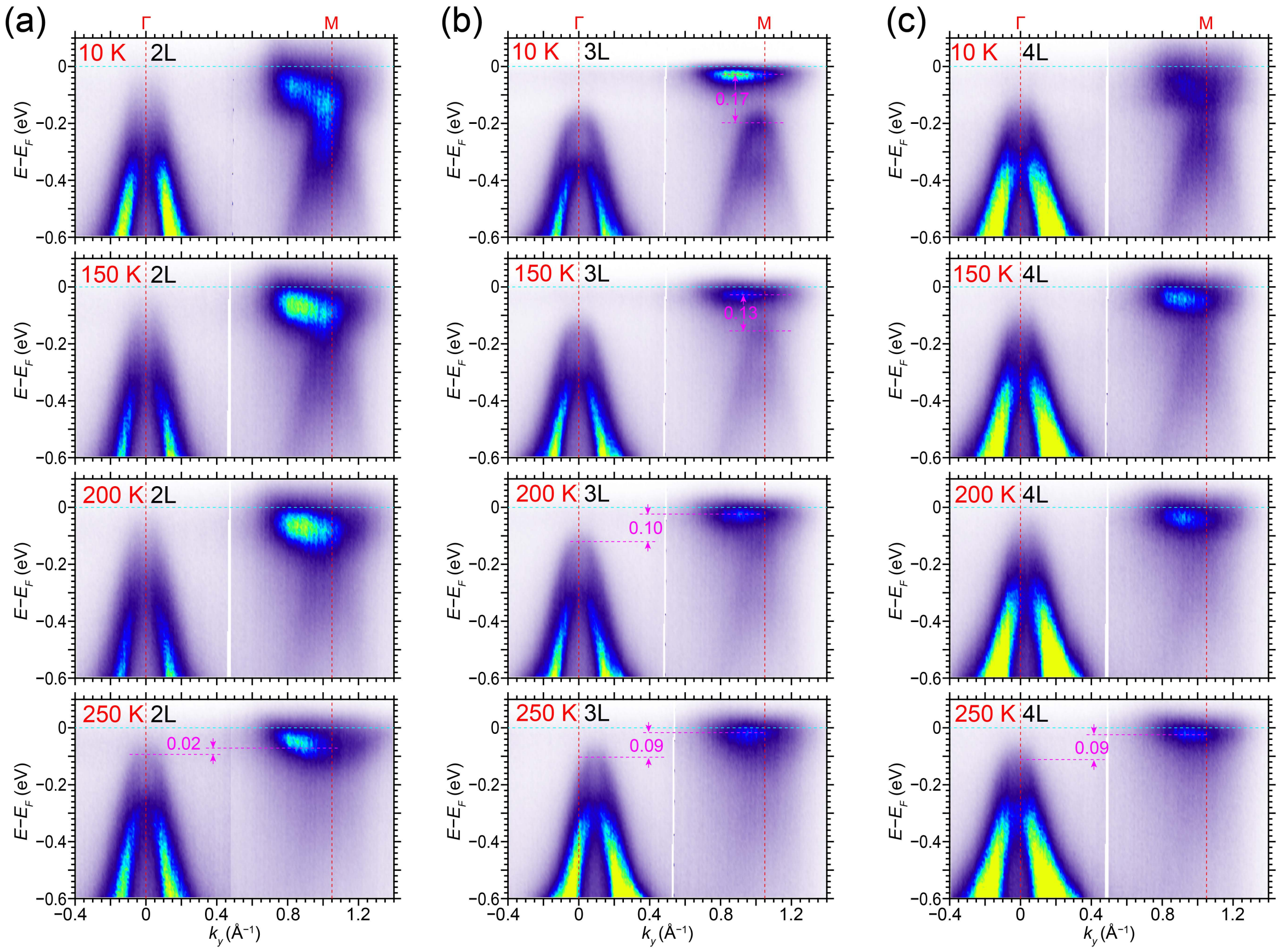

**Fig. S6. (a)-(c)** VT-ARPES spectra on the VB folding along the Γ-M direction of (a) 2L, (b) 3L and (c) 4L $TiSe_2$ films, respectively.

## G. Calculation method on the excitonic state in even-layers $TiSe_2$

In the bilayer $TiSe_2$, the ARPES spectrum exhibits a clear depletion of spectral weight near the top of the valence band, in sharp contrast to the monolayer case. We attribute this suppression of the valence-band spectral weight to the formation of excitons. Meanwhile, the spectral feature associated with the conduction band becomes significantly broadened and less well defined. As will be shown in the following, this broadening has several distinct physical origins, i.e., the renormalization effect between the conduction band electrons, the exciton formation, as well as the finite electron life time magnified by density imbalance.

### G1. Renormalization effect between conduction band electrons

To capture the low-energy band structure of bilayer $TiSe_2$ near the Fermi level, we start from a bare continuum model around the Γ point. This model describes the valence and conduction bands of the two layers, with the Fermi level fixed at $E_F = 0$, as shown by the dispersion along $k_x = 0$ direction in Fig. S7.

Since the depletion of the spectral weight near the top of the valence-band and the broadening of the conduction band are not observed in monolayer $TiSe_2$, their appearance in bilayers indicates that these features may originate from interlayer interactions. Thus, we consider the screened interlayer Coulomb interaction. The interlayer bare Coulomb interaction is given by $V_{\text{inter}}(q) = V_{\text{intra}}(q)e^{-qd}$, where $d$ is the layer separation. Within the random phase approximation (RPA), the dynamically screened interlayer interaction is written as $V(\boldsymbol{q},\omega)=\frac{V_{\text{inter}}(q)}{\epsilon(\boldsymbol{q},\omega)}$ , where the dielectric function of the coupled-layer system is given by Ref. [8], i.e.,

$$\epsilon(\boldsymbol{q},\omega)=1-V_{\text{inter}}(q)\Pi(\boldsymbol{q},\omega) \tag{S4}$$

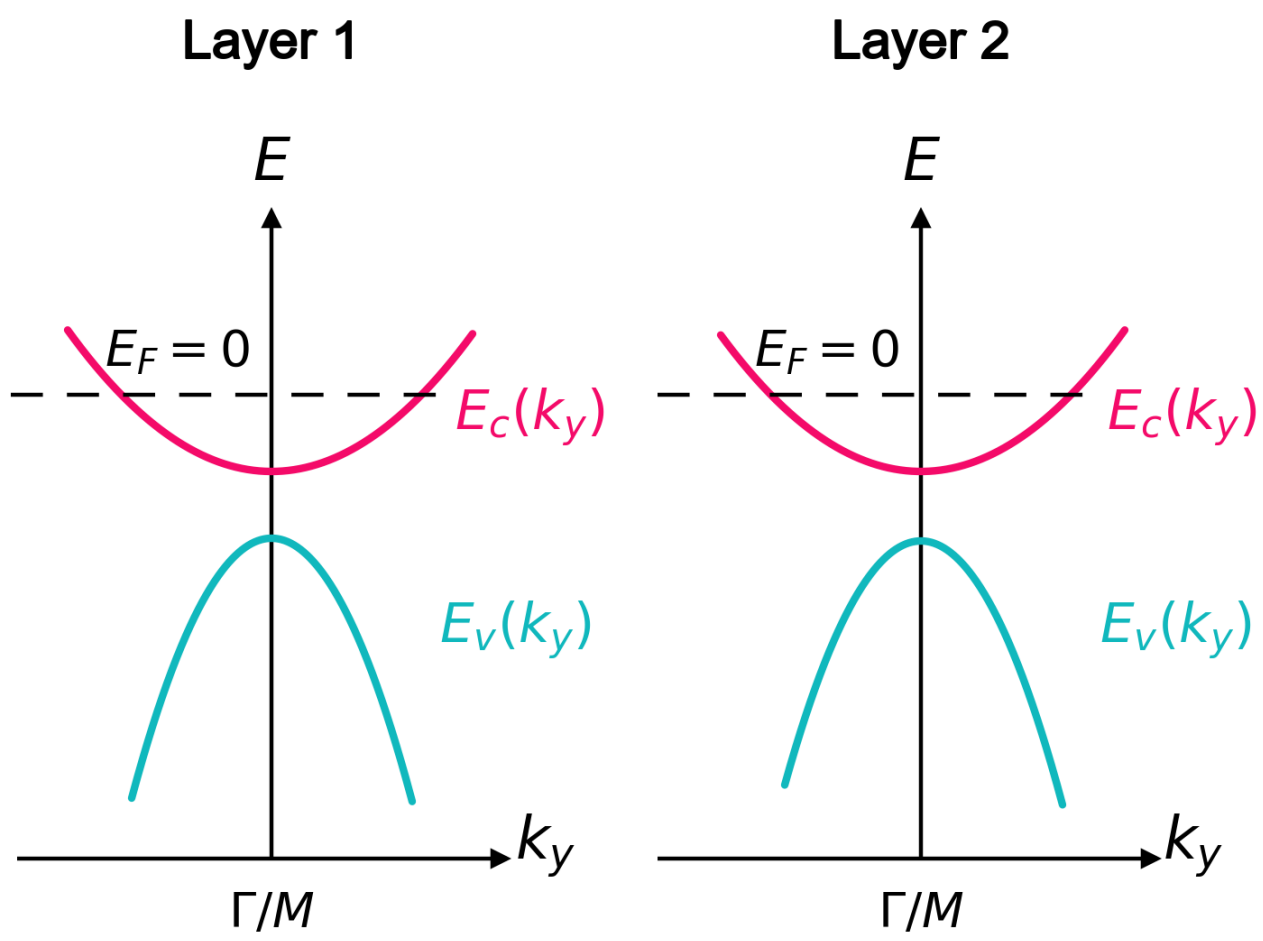

**Fig. S7.** Schematic band dispersions of bilayer $TiSe_2$ in the continuum model. The conduction and valence bands of the two layers are plotted along the $k_y$ direction ($k_x = 0$) near the Γ/M point. The band dispersions are given by $E_c(k_y) = k^2 - 0.1$ and $E_v(k_y) = -2k^2 - 0.2$, respectively. The dashed line denotes the Fermi level, $E_F = 0$.

Here $\Pi(\boldsymbol{q},\omega)$ is the Lindhard function. In addition, the bare intralayer Coulomb interaction is given by $V_{\text{intra}}(q)=\frac{2\pi e^2}{\kappa q}$, where $\kappa$ is the dielectric constant.

We first discuss the possible origin of the broadening of the conduction band observed by ARPES. In bilayer $TiSe_2$, the conduction-band electrons in the two layers interact with each other through the interlayer Coulomb interaction. Therefore, we consider the self-energy correction to the conduction-band electrons in the first layer induced by the conduction-band electrons in the second layer. The renormalization by the valence band electrons in the second layer is negligible, because the valence band is fully occupied and the relevant microscopic processes are strongly suppressed by Pauli blocking. The corresponding self-energy can be decomposed into the Hartree and Fock contributions, as shown diagrammatically in Fig. S8.

In the retarded form, the interlayer conduction-band Lindhard function is

$$\Pi^R(\boldsymbol{q},\nu)=\int\frac{d^2\boldsymbol{k}}{(2\pi)^2}\frac{n_F(E_{1,c}(\boldsymbol{k}))-n_F(E_{2,c}(\boldsymbol{k}+\boldsymbol{q}))}{\nu+i\eta+E_{1,c}(\boldsymbol{k})-E_{2,c}(\boldsymbol{k}+\boldsymbol{q})} \tag{S5}$$

where $n_F(x)=1/(e^{x/(k_BT)}+1)$ is the Fermi-Dirac distribution, $\eta\to0^+$ is a positive infinitesimal, and $E_{1,c}(\boldsymbol{k})$ and $E_{2,c}(\boldsymbol{k}+\boldsymbol{q})$ denote the conduction-band energy dispersion in the first and second layers, respectively. In the following numerical calculations, we set $\kappa = 5$. We use natural units for the Coulomb interaction and set the electron charge to be $e = 1$. Moreover, lengths are measured in units of the in-plane lattice constant $\boldsymbol{a}$, so that $\boldsymbol{a} = 1$, while momenta are measured in units of $\boldsymbol{a}^{-1}$. For bulk 1T-$TiSe_2$, the experimental lattice parameters are $\boldsymbol{a} = \boldsymbol{b} = 3.540$ Å and $\boldsymbol{c} = 6.008$ Å [9]. The parameter $\boldsymbol{c}$ is the out-of-plane lattice constant. Since the 1T unit cell contains one Se-Ti-Se layer along the $\boldsymbol{c}$ axis, we approximate the layer separation by $\boldsymbol{c}$. Therefore, in lattice units, $d = \frac{\boldsymbol{c}}{\boldsymbol{a}} = \frac{6.008}{3.540} \simeq 1.70$.

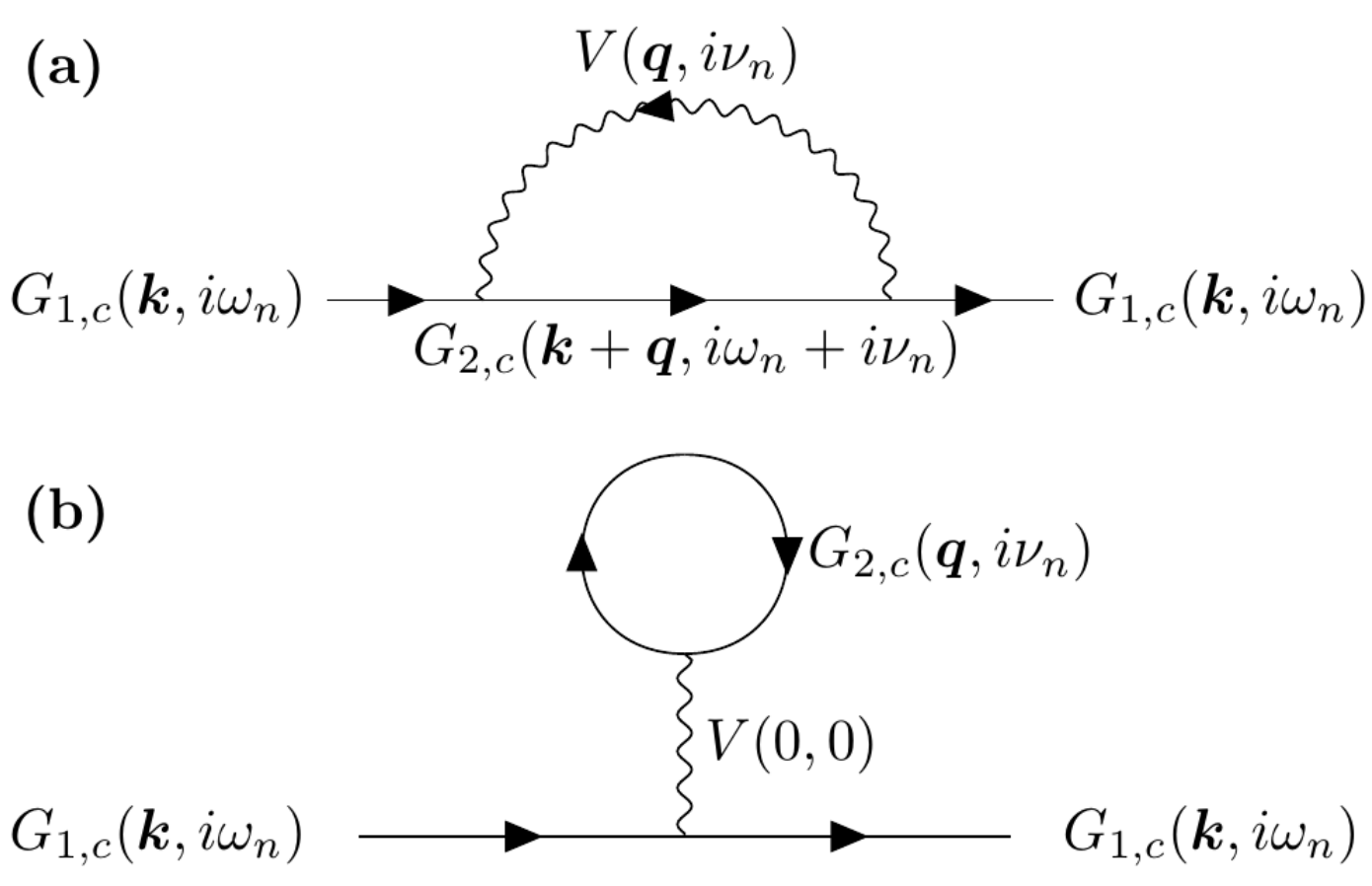

**Fig. S8.** Self-energy diagrams. **(a)** Fock self-energy diagram. **(b)** Hartree self-energy diagram.

The retarded Green's function of the conduction band in Layer 1 is given by

$$G^R_{1,c}(\boldsymbol{k},\omega) = \frac{1}{\omega + i\eta - E_{1,c}(\boldsymbol{k}) - \Sigma^R_{1,cc}(\boldsymbol{k},\omega)} \tag{S6}$$

where $\Sigma^R_{1,cc}(\boldsymbol{k},\omega)$ denotes the retarded self-energy correction arising from the interlayer interaction with the conduction-band electrons in Layer 2. The corresponding spectral function is given by

$$A_{1,c}(\boldsymbol{k},\omega) = -2\mathrm{Im}G^R_{1,c}(\boldsymbol{k},\omega) \tag{S7}$$

A finite imaginary part of $\Sigma^R_{1,cc}(\boldsymbol{k},\omega)$ gives rise to a finite quasiparticle lifetime, which leads to the broadening of the conduction-band spectral feature.

The Hartree self-energy is obtained as

$$\begin{aligned}
\Sigma^H_{1,c}(\boldsymbol{p}, ip_n) &= -\frac{1}{\beta S}\sum_{\boldsymbol{p}}\sum_{ip_n}[-V(0,0)]\, G^0_{2,c}(\boldsymbol{p}, ip_n)e^{ip_n\eta} \\
&= \frac{V(0,0)}{\beta}\int \frac{d^2\boldsymbol{p}}{(2\pi)^2}\sum_{ip_n}\frac{e^{ip_n\eta}}{ip_n - E_{2,c}(\boldsymbol{p})} \\
&= V(0,0)\int \frac{d^2\boldsymbol{p}}{(2\pi)^2} n_F(E_{2,c}(\boldsymbol{p})) = V(0,0)\frac{N_{2,c}}{S}
\end{aligned} \tag{S8}$$

where $N_{2,c}$ denotes the number of electrons in the conduction band of the second layer, and $S$ is the system area. Since the Hartree self-energy is independent of both momentum and frequency, its retarded form is obtained simply by the analytic continuation $i\omega_n \to \omega + i0^+$,

$$\Sigma_{1,c}^{R,H}(\boldsymbol{k},\omega) = V(0,0)\frac{N_{2,c}}{S}. \tag{S9}$$

Note that the static polarizability is given by $\Pi(0,0) = \int \frac{d^2\boldsymbol{k}}{(2\pi)^2}\frac{\partial n_F(E_c(\boldsymbol{k}))}{\partial E_c(\boldsymbol{k})}$, and the electron density in the conduction band of the second layer is $\frac{N_{2,c}}{S} = \int \frac{d^2\boldsymbol{k}}{(2\pi)^2} n_F(E_{2,c}(\boldsymbol{k}))$, from which we can calculate $\Sigma_{1,c}^{R,H}(\boldsymbol{k},\omega) = 0.1$ eV. Therefore, the Hartree term only produces a constant energy shift of the conduction band in Layer 1 and does not contribute to the spectral broadening. The uniform Hartree contribution is assumed to be canceled by the positively charged ionic background and therefore can be omitted. The imaginary part of Fock self-energy is written as [10]

$$\begin{aligned}\mathrm{Im}\Sigma_{1,c}^{F}(\boldsymbol{k},\epsilon) = \int \frac{d^2\boldsymbol{q}}{(2\pi)^2}\big[n_B(E_{2,c}(\boldsymbol{k}+\boldsymbol{q})-\epsilon) + n_F(E_{2,c}(\boldsymbol{k} \\ + \boldsymbol{q}))\big]\mathrm{Im}\left[\frac{V(q)}{\epsilon\big(\boldsymbol{q}, E_{2,c}(\boldsymbol{k}+\boldsymbol{q}) - \epsilon + i\eta\big)}\right]\end{aligned} \tag{S10}$$

where $n_B(x) = \frac{1}{e^{x/(k_BT)}-1}$ is the bosonic distribution. The real part of the Fock self-energy can then be obtained from the Kramers-Kronig relation,

$$\mathrm{Re}\Sigma_{1,c}^{F}(\boldsymbol{k},\epsilon) = \frac{1}{\pi}\mathcal{P}\int_{-\infty}^{+\infty} d\,\epsilon' \frac{\mathrm{Im}\Sigma_{1,c}^{F}(\boldsymbol{k},\epsilon')}{\epsilon' - \epsilon} \tag{S11}$$

where $\mathcal{P}$ denotes the Cauchy principal value. To improve the numerical convergence, we use the subtracted Kramers-Kronig relation. Starting from the conventional relation and subtracting its value at $\epsilon = 0$, we obtain

$$\mathrm{Re}\Sigma_{1,c}^{F}(\boldsymbol{k},\epsilon) - \mathrm{Re}\Sigma_{1,c}^{F}(\boldsymbol{k},0) = \frac{\epsilon}{\pi}\mathcal{P}\int_{-\infty}^{+\infty} d\,\epsilon' \frac{\mathrm{Im}\Sigma_{1,c}^{F}(\boldsymbol{k},\epsilon')}{\epsilon'(\epsilon' - \epsilon)} \tag{S12}$$

To improve the numerical convergence, we first evaluate the energy-dependent part of the real self-energy, $\mathrm{Re}\Sigma_{1,c}^{F}(\boldsymbol{k},0) - \mathrm{Re}\Sigma_{1,c}^{F}(\boldsymbol{k},0)$, using the subtracted Kramers-Kronig relation in Eq. (S12). The zero-energy contribution$\mathrm{Re}\Sigma_{1,c}^{F}(\boldsymbol{k},0)$ is evaluated separately. The full real part of the self-energy is then reconstructed by combining these two contributions. In the numerical calculation, we set $T = 10$ K, which corresponds to $\beta = 1160\ \mathrm{eV}^{-1}$. Eq. (S10) leads to the broadening of the conduction band. This is one of the factors accounting for the experimentally broadening. As will be clear below, another factor is attributed to the finite damping rate induced by density imbalance, which leaves a finite population of unpaired conduction-band electrons coexisting with the interlayer excitons

### G2. Excitonic pairing channel

To characterize the interlayer particle-hole excitation between the valence band of the first layer and the conduction band of the second layer, we introduce the corresponding bare inter-band polarization function, which, in the retarded form, reads as,

$$\Pi^R_{0,v_1c_2}(\mathbf{q},\omega)=\int\frac{d^2\mathbf{k}}{(2\pi)^2}\frac{n_F\big(E_{1,v}(\mathbf{k})\big)-n_F\big(E_{2,c}(\mathbf{k}+\mathbf{q})\big)}{\omega+i\eta+E_{1,v}(\mathbf{k})-E_{2,c}(\mathbf{k}+\mathbf{q})} \tag{S13}$$

where $E_{1,v}(\mathbf{k})$ is the valence-band dispersion in the first layer, $E_{2,c}(\mathbf{k}+\mathbf{q})$ is the conduction-band dispersion in the second layer, and $\eta\to 0^+$. This polarization describes the virtual excitation of an electron from the occupied valence band of the first layer to the conduction band of the second layer, leaving a hole in the first layer. Therefore, $\Pi^R_{0,v_1c_2}(\mathbf{q},\omega)$ represents the bare interlayer interband response in the particle-hole channel.

To analyze the tendency toward an interlayer excitonic instability, we focus on the particle-hole channel connecting the valence band of the first layer and the conduction band of the second layer. In the static limit, the bare particle-hole susceptibility in the excitonic channel as is

$$\chi^0_{\mathrm{eh}}(\mathbf{q})=-\mathrm{Re}\,\Pi^R_{0,v_1c_2}(\mathbf{q},0) \tag{S14}$$

Explicitly, one has

$$\chi^0_{\mathrm{eh}}(\mathbf{q})=\int\frac{d^2\mathbf{k}}{(2\pi)^2}\frac{n_F\big(E_{1,v}(\mathbf{k})\big)-n_F\big(E_{2,c}(\mathbf{k}+\mathbf{q})\big)}{E_{2,c}(\mathbf{k}+\mathbf{q})-E_{1,v}(\mathbf{k})} \tag{S15}$$

The instability condition is therefore given by,

$$1-V_{\mathrm{inter}}(|\mathbf{Q}|)\chi^0_{\mathrm{eh}}(\mathbf{Q})=0 \tag{S16}$$

or equivalently,

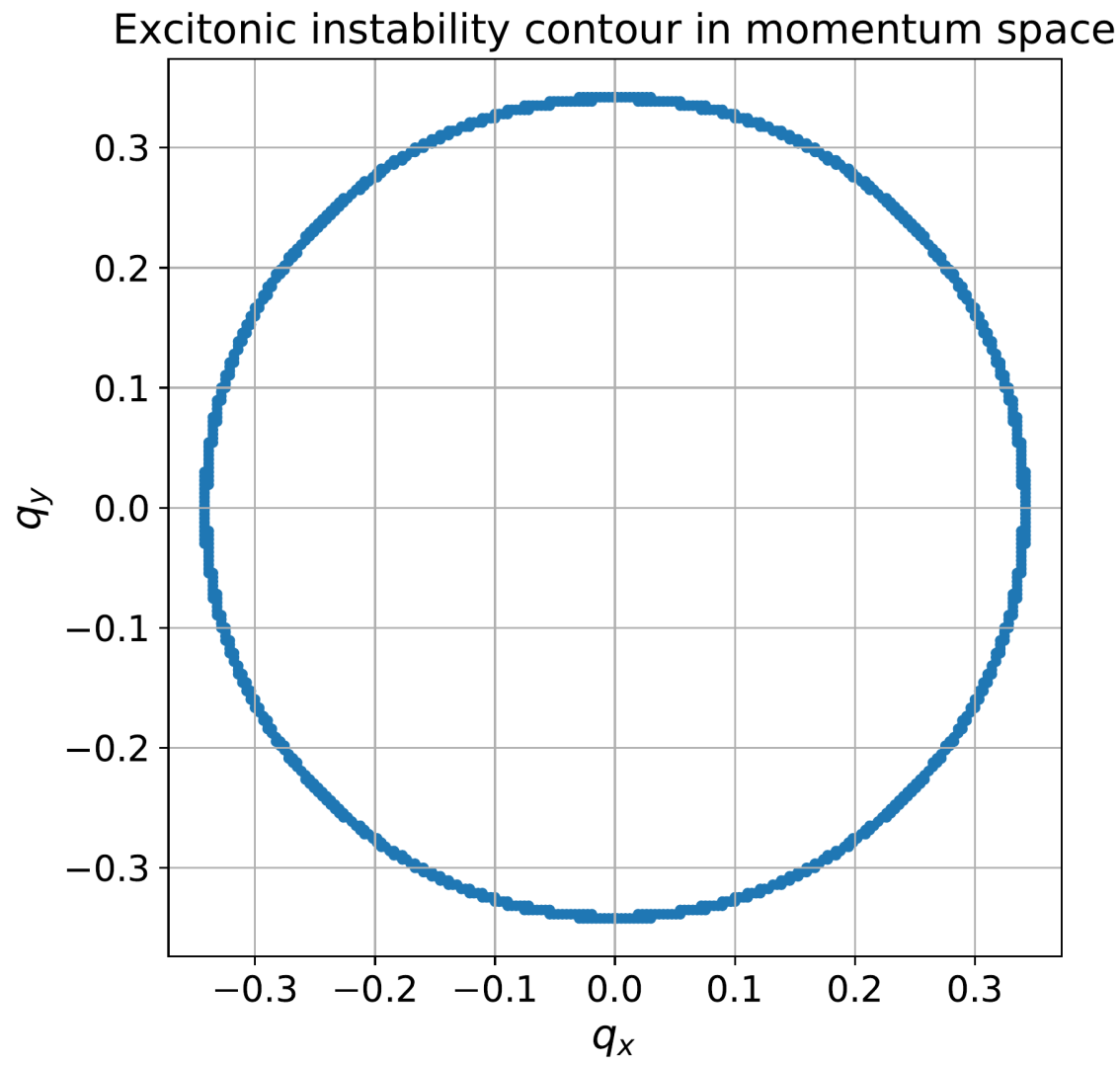

**Fig. S9.** Zero contour of the excitonic instability denominator $1 + V_{\text{inter}}(q)\text{Re}\Pi^R_{0,v_1c_2}(\mathbf{q}, 0)$. The nearly circular contour indicates that the instability occurs at a finite ordering wave vector $|\mathbf{Q}|$. The radius of the contour is close to the Fermi momentum of the conduction band, suggesting that the interlayer particle-hole excitation mainly connects the valence-band top to the conduction-band Fermi surface.

$$1 + V_{\text{inter}}(|\mathbf{Q}|)\text{Re}\Pi^R_{0,v_1c_2}(\mathbf{Q}, 0) = 0 \tag{S17}$$

where $\mathbf{Q}$ is the interlayer ordering wave vector.

Figure S9 shows all the wave vectors satisfying the above excitonic instability condition, $1 + V_{\text{inter}}(q)\text{Re}\Pi^R_{0,v_1c_2}(\mathbf{q}, 0) = 0$. The resulting zero contour is nearly circular, reflecting the rotational symmetry of the simplified continuum model. Its radius is approximately equal to the Fermi momentum of the conduction band. Therefore, the dominant interlayer particle-hole channel connects states near the valence-band top in the first layer to states around the conduction-band Fermi surface in the second layer. This indicates possible depletion of the ARPES spectral weight near the valence-band maximum: electrons near the valence-band top are strongly involved in the interlayer excitonic channel and their single-particle spectral weight is consequently suppressed.

### G3. LO-type interlayer excitonic reconstruction

Once the instability condition is satisfied, the valence-band Green's function in the first layer can be strongly renormalized by its coupling to the conduction band in the second layer. In the ideal continuum model, the system is assumed to be isotropic, and Eq.(S19) is simultaneously satisfied by all wave vectors belong to the momentum loop in Fig. S9. This leads to strong quantum fluctuations and possible excitonic topological orders with long-range quantum entanglement. However, in real materials, the crystalline anisotropy lifts the continuous directional degeneracy and could favor exciton condensation along a discrete set of preferred crystallographic directions. This results in Fulde-Ferrell (FF) or Larkin-Ovchinnikov (LO) type excitonic condensates. Taking the LO state as an example, the excitons coherently condensate at two opposite wave vectors, $+\boldsymbol{Q}$ and $-\boldsymbol{Q}$. Thus, we define the corresponding interlayer excitonic order parameters as

$$\Delta_{\pm\mathbf{Q}}(\mathbf{k}) = V_{\text{inter}}(\mathbf{Q})\langle c_{2,c}^{\dagger}(\mathbf{k} \pm \mathbf{Q})c_{1,v}(\mathbf{k})\rangle \tag{S18}$$

The real-space excitonic order parameter is therefore given by

$$\Delta(\boldsymbol{r}) = \Delta_{+\mathbf{Q}}e^{i\mathbf{Q}\cdot\boldsymbol{r}} + \Delta_{-\mathbf{Q}}e^{-i\mathbf{Q}\cdot\boldsymbol{r}} \tag{S19}$$

For an equal-amplitude LO state with $\Delta_{+\boldsymbol{Q}} = \Delta_{-\boldsymbol{Q}} \equiv \Delta_{\boldsymbol{Q}}$, Eq. (S19) reduces to

$$\Delta(\boldsymbol{r}) = 2\Delta_{\mathbf{Q}}\cos(\mathbf{Q}\cdot\boldsymbol{r}) \tag{S20}$$

which describes an amplitude-modulated interlayer excitonic state.

In the continuum description, the valence-band maximum is located near $\boldsymbol{k} = 0$, whereas the conduction band forms an approximately circular Fermi surface with radius $k_F$. The instability contour satisfies $|\boldsymbol{Q}| \simeq k_F$, such that the valence-band states near $\boldsymbol{k} = 0$ are resonantly coupled to conduction-band states near both $\boldsymbol{k} + \boldsymbol{Q}$ and $\boldsymbol{k} - \boldsymbol{Q}$. We parameterize the ordering vector as

$$\mathbf{Q} = \mathbf{Q}(\cos\,\theta_{\mathbf{Q}}, \sin\,\theta_{\mathbf{Q}}) \tag{S21}$$

For the calculation presented below, we take $Q = \sqrt{0.1}$ and $\theta_Q = 0$.

To describe the low-energy spectral reconstruction near the valence-band maximum, we retain the first-layer valence state $c_{1,v}(\boldsymbol{k})$ and the two second-layer conduction states $c_{2,c}(\boldsymbol{k}+\boldsymbol{Q})$ and $c_{2,c}(\boldsymbol{k}-\boldsymbol{Q})$ for small $\boldsymbol{k}$. The approximate mean-field Hamiltonian is then given by,

$$H_{LO} = \sum_{\mathbf{k}} \begin{pmatrix} c_{1,v}^{\dagger}(\mathbf{k}) & c_{2,c}^{\dagger}(\mathbf{k}+\mathbf{Q}) & c_{2,c}^{\dagger}(\mathbf{k}-\mathbf{Q}) \end{pmatrix} \begin{pmatrix} E_{1,v}(\mathbf{k}) & \Delta_{+\mathbf{Q}}^{*}(\mathbf{k}) & \Delta_{-\mathbf{Q}}^{*}(\mathbf{k}) \\ \Delta_{+\mathbf{Q}}(\mathbf{k}) & E_{2,c}(\mathbf{k}+\mathbf{Q}) & 0 \\ \Delta_{-\mathbf{Q}}(\mathbf{k}) & 0 & E_{2,c}(\mathbf{k}-\mathbf{Q}) \end{pmatrix} \begin{pmatrix} c_{1,v}(\mathbf{k}) \\ c_{2,c}(\mathbf{k}+\mathbf{Q}) \\ c_{2,c}(\mathbf{k}-\mathbf{Q}) \end{pmatrix} \tag{S22}$$

The corresponding retarded Green's function is a $3 \times 3$ matrix,

$$\hat{\mathrm{G}}^{R}_{\mathrm{LO}}(\mathbf{k},\omega) = \left[(\omega + i\eta)\hat{\mathrm{I}} - \hat{\mathrm{H}}_{\mathrm{LO}}(\mathbf{k})\right]^{-1} \tag{S23}$$

Explicitly,

$$\hat{\mathrm{G}}^{R}_{\mathrm{LO}}(\mathbf{k},\omega) = \left[\begin{pmatrix} \omega + i\eta - E_{1,v}(\mathbf{k}) & -\Delta^{*}_{+\mathbf{Q}}(\mathbf{k}) & -\Delta^{*}_{-\mathbf{Q}}(\mathbf{k}) \\ -\Delta_{+\mathbf{Q}}(\mathbf{k}) & \omega + i\eta - E_{2,c}(\mathbf{k}+\mathbf{Q}) & 0 \\ -\Delta_{-\mathbf{Q}}(\mathbf{k}) & 0 & \omega + i\eta - E_{2,c}(\mathbf{k}-\mathbf{Q}) \end{pmatrix}\right]^{-1} \tag{S24}$$

Taking the (1,1) component of Eq. (S24), we obtain the Green's function of the first-layer valence band as,

$$G^{R}_{1,v}(\mathbf{k},\omega) = \frac{1}{\omega + i\eta - E_{1,v}(\mathbf{k}) - \Sigma^{R}_{1,v,\mathrm{LO}}(\mathbf{k},\omega)} \tag{S25}$$

where the LO excitonic channel generates the valence-band self-energy,

$$\Sigma^{R}_{1,v,\mathrm{LO}}(\mathbf{k},\omega) = \frac{|\Delta_{+\mathbf{Q}}(\mathbf{k})|^2}{\omega + i\Gamma - E_{2,c}(\mathbf{k}+\mathbf{Q})} + \frac{|\Delta_{-\mathbf{Q}}(\mathbf{k})|^2}{\omega + i\Gamma - E_{2,c}(\mathbf{k}-\mathbf{Q})} \tag{S26}$$

and the corresponding valence-band spectral function is obtained via

$$A_{1,v}(\mathbf{k},\omega) = -2\mathrm{Im}G^{R}_{1,v}(\mathbf{k},\omega) \tag{S27}$$

The two terms in Eq. (S26) describe virtual processes in which an electron near the valence-band top in the first layer is scattered into conduction-band states of the second layer with momenta $\boldsymbol{k}+\boldsymbol{Q}$ and $\boldsymbol{k}-\boldsymbol{Q}$, respectively, and then returns back to the original valence-band state. To account for the finite lifetime of the intermediate interlayer excitonic channel, we introduce a phenomenological damping rate $\Gamma$ in the two conduction-band propagators (as analyzed in the next section). Here, $\Gamma$ describes the finite lifetime of the conduction-band states, which broadens the spectral function and generates a depletion near top of the valence-band. In our calculations, an effective total excitonic order is introduced as $|\Delta_{+\mathbf{Q}}|^2 + |\Delta_{-\mathbf{Q}}|^2 = |\Delta_{eff}|^2$, and $\Delta_{+\mathbf{Q}} = \Delta_{-\mathbf{Q}} = \frac{\Delta_{eff}}{\sqrt{2}}$, and $\Delta_{\mathrm{eff}} = 0.2$ eV and $\Gamma = 0.2$ eV.

We next consider the influence of the LO-type interlayer excitonic order on the conduction-band Green's function. A conduction-band state with momentum $\boldsymbol{k}$ is coupled to the valence-band states with momentum $\boldsymbol{k}-\boldsymbol{Q}$ and $\boldsymbol{k}+\boldsymbol{Q}$. This leads to the following retarded self-energy of the conduction band,

$$\Sigma^{R,\mathrm{ex}}_{1,c,\mathrm{LO}}(\mathbf{k},\omega) = \frac{|\Delta_{+\mathbf{Q}}|^2}{\omega + i\Gamma - E_{2,v}(\mathbf{k}-\mathbf{Q})} + \frac{|\Delta_{-\mathbf{Q}}|^2}{\omega + i\Gamma - E_{2,v}(\mathbf{k}+\mathbf{Q})} \tag{S28}$$

where $\Gamma$ denotes the phenomenological broadening of the intermediate valence-band states.

Taking into account both the self-energy $\Sigma^R_{1,cc}(\boldsymbol{k},\omega)$ induced by the interaction between the conduction-band electrons in the two layers (in Eq.(S6)) and the LO excitonic contribution, the retarded Green's function of the conduction-band is obtained as,

$$G^R_{1,c}(\mathbf{k},\omega) = \frac{1}{\omega + i\Gamma_c - E_{1,c}(\mathbf{k}) - \Sigma^R_{1,cc}(\mathbf{k},\omega) - \Sigma^{R,\mathrm{ex}}_{1,c,\mathrm{LO}}(\mathbf{k},\omega)} \tag{S29}$$

The corresponding conduction-band spectral function is

$$A_{1,c}(\mathbf{k},\omega) = -2\mathrm{Im}G^R_{1,c}(\mathbf{k},\omega) \tag{S30}$$

Fig. 4(e) in the main text is obtained using Eq.(S30) and plot along the $k_x = 0$ momentum cut.

### G4. The finite lifetimes of conduction-band electrons and excitons

We now discuss the physical origins of the finite lifetime, or equivalently the spectral broadening, parametrized by the parameters $\eta_c$ and $\Gamma$ above. In the present system, the bottom of the conduction band is filled. Being away from the charge neutral point results in density imbalance between electron and holes. The density imbalance leaves a finite population of unpaired conduction-band electrons coexisting with the interlayer excitons [11]. Scattering between the excitons and these excess electrons provides an intrinsic damping channel, resulting in a finite lifetime of the interlayer excitons [12]. This results in a finite damping rate $\Gamma$ in the excitonic self-energy. More precisely, electron-exciton scattering may involve exchange processes in which an excess conduction-band electron exchanges with the electron constituent of an exciton. Such scattering processes broaden both the conduction-electron and exciton spectral responses, leading to finite lifetimes for both species. These effects are captured by the conduction-band broadening parameter $\eta_c$ and the exciton damping rate $\Gamma$, respectively.

As illustrated in Fig. S10(a), the excess conduction-band electrons interact with the excitons through,

$$H_{\mathrm{e-ex}} = g \sum_{\mathbf{k},\boldsymbol{q}} c^\dagger_{\mathbf{k}+\boldsymbol{q}} c_{\mathbf{k}} \left(b_{\boldsymbol{q}} + b^\dagger_{-\boldsymbol{q}}\right) \tag{S31}$$

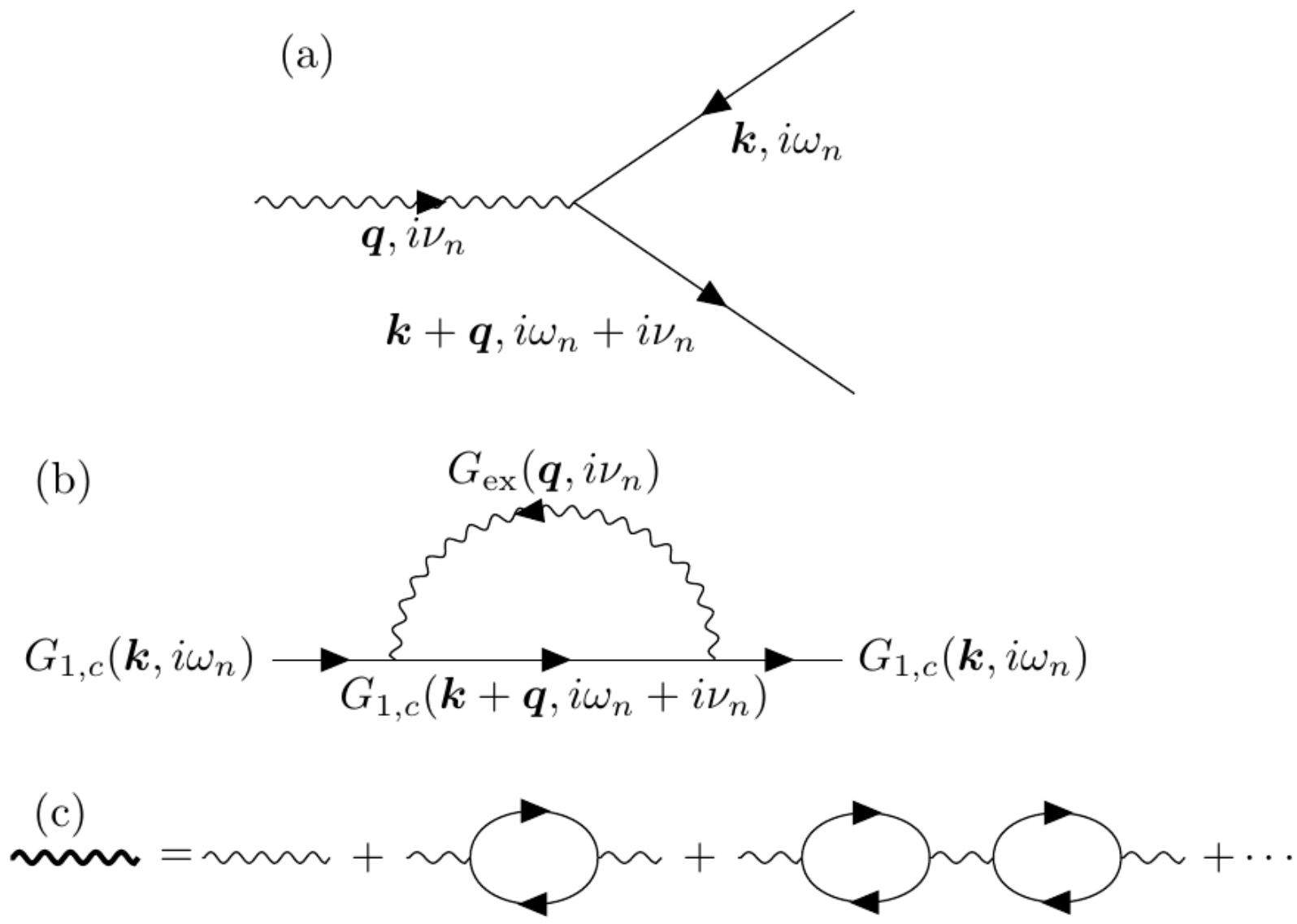

**Fig. S10.** Feynman diagrams for the electron-exciton coupling and the corresponding self-energy corrections. **(a)** Electron-exciton interaction vertex. **(b)** Self-energy correction to a conduction-band electron induced by exciton exchange. **(c)** Dyson series for the dressed exciton propagator generated by electron particle-hole polarization bubbles.

where $c_{\mathbf{k}}$ and $b_{\boldsymbol{q}}$ denote the annihilation operators of a conduction-band electron and an excitonic bosonic mode, respectively. As shown in Fig. S10(b), the exchange of an excitonic mode generates a self-energy correction to the conduction-band electron. At the one-loop order, the electronic self-energy is given by

$$\Sigma_{1,c}^{\text{e-ex}}(\mathbf{k}, i\omega_n) = \frac{g^2}{\beta} \sum_{\boldsymbol{q}, i\nu_n} G_{\text{ex}}(\boldsymbol{q}, i\nu_n) G_{1,c}^0(\mathbf{k} + \boldsymbol{q}, i\omega_n + i\nu_n) \tag{S32}$$

where $G_{1,c}^0$ is the bare Green's function of the conduction-band electrons and $G_{\text{ex}}$ is the exciton propagator. The corresponding dressed electronic Green's function takes the form

$$G_{1,c}^{R,\text{e-ex}}(\mathbf{k}, \omega) = \frac{1}{\omega - E_{1,c}(\mathbf{k}) - \Sigma_{1,c}^{R,\text{e-ex}}(\mathbf{k}, \omega)} \tag{S33}$$

When the electron can emit or absorb an excitonic excitation while satisfying energy and momentum conservation, the electronic self-energy acquires a finite imaginary part $\eta_c$. Therefore, scattering from excitonic fluctuations broadens the conduction-electron spectral function and renders the conduction-band quasiparticles finite-lived. The real part of the self-energy primarily produces a smooth renormalization of the exciton energy and dispersion, and we do not retain it explicitly.

Conversely, the excess conduction-band electrons also renormalize the excitonic propagator. As illustrated in Fig. S10(c), the electron particle-hole polarization produces an exciton self-energy

$$\Sigma_{\mathrm{ex}}(\boldsymbol{q}, i\nu_n) = g^2 \Pi_{\mathrm{e}}(\boldsymbol{q}, i\nu_n) \tag{S34}$$

where

$$\Pi_{\mathrm{e}}(\boldsymbol{q}, i\nu_n) = -\frac{1}{\beta} \sum_{\mathbf{k}, i\omega_n} G^0_{1,c}(\mathbf{k}, i\omega_n) G^0_{1,c}(\mathbf{k}+\boldsymbol{q}, i\omega_n + i\nu_n) \tag{S35}$$

is the polarization function of the excess electrons. Summing the Dyson series in Fig. S10(c), one obtains the fully dressed exciton Green's function

$$\begin{aligned} G_{\mathrm{ex}} &= G^0_{\mathrm{ex}} + G^0_{\mathrm{ex}} \Sigma_{\mathrm{ex}} G^0_{\mathrm{ex}} + G^0_{\mathrm{ex}} \Sigma_{\mathrm{ex}} G^0_{\mathrm{ex}} \Sigma_{\mathrm{ex}} G^0_{\mathrm{ex}} + \cdots \\ &= \frac{1}{(G^0_{\mathrm{ex}})^{-1} - \Sigma_{\mathrm{ex}}} \end{aligned} \tag{S36}$$

After the analytic continuation $i\nu_n \to \nu + i0^+$, the retarded exciton Green's function is

$$G^R_{\mathrm{ex}}(\boldsymbol{q}, \nu) = \frac{1}{\left[G^{0,R}_{\mathrm{ex}}(\boldsymbol{q}, \nu)\right]^{-1} - \Sigma^R_{\mathrm{ex}}(\boldsymbol{q}, \nu)} \tag{S37}$$

The retarded exciton self-energy can be decomposed as

$$\Sigma^R_{\mathrm{ex}}(\boldsymbol{q}, \nu) = \mathrm{Re}\Sigma^R_{\mathrm{ex}}(\boldsymbol{q}, \nu) - i\Gamma(\boldsymbol{q}, \nu) \tag{S38}$$

where

$$\Gamma(\boldsymbol{q}, \nu) = -\mathrm{Im}\Sigma^R_{\mathrm{ex}}(\boldsymbol{q}, \nu) \geq 0 \tag{S39}$$

is the damping rate induced by the electron particle-hole polarization. The dressed exciton Green's function is therefore given by

$$G^R_{\mathrm{ex}}(\boldsymbol{q}, \nu) = \frac{1}{\left[G^{0,R}_{\mathrm{ex}}(\boldsymbol{q}, \nu)\right]^{-1} - \mathrm{Re}\Sigma^R_{\mathrm{ex}}(\boldsymbol{q}, \nu) + i\Gamma(\boldsymbol{q}, \nu)} \tag{S40}$$

The real part of the self-energy renormalizes the exciton dispersion, whereas its imaginary part broadens the excitonic spectral peak and gives rise to a finite exciton lifetime. In general, the damping rate $\Gamma(\boldsymbol{q}, \nu)$ depends on both momentum and frequency. Since a microscopic evaluation of the electron-exciton scattering rate is beyond the scope of the present analysis, we approximate $\Gamma(\boldsymbol{q}, \nu)$ by a constant phenomenological value and take $\Gamma = 0.2$ eV over the relevant low-energy window to account for the ARPES data. This relatively large damping rate describes a short-lived and strongly broadened excitonic mode, consistent with the coexistence of interlayer excitons and excess conduction-band electrons, as described by Ref.[11]. The real part of the exciton self-energy primarily produces a smooth renormalization of the exciton energy and dispersion. Since the exciton dispersion used in the present phenomenological model is regarded as an effective renormalized dispersion, we do not retain it explicitly.